\documentclass[lettersize,journal]{IEEEtran}
\usepackage{amsmath,amsfonts}
\usepackage{amssymb}
\usepackage{algorithmic}
\usepackage{algorithm}
\usepackage{array}
\usepackage{booktabs} 
\usepackage[caption=false,font=normalsize,labelfont=sf,textfont=sf]{subfig}
\usepackage{textcomp}
\usepackage{stfloats}
\usepackage{url}
\usepackage{verbatim}
\usepackage{graphicx}
\usepackage{cite}
\usepackage{xcolor}
\usepackage{graphicx}

\begin{document}

\title{Enhancing WiFi CSI Fingerprinting:\\ A Deep Auxiliary Learning Approach}

\author{Yong Huang,~\IEEEmembership{Member,~IEEE}, Wenjing Wang, Dalong Zhang, Junjie Wang, Chen Chen, Yan Cao, \\ Wei Wang,~\IEEEmembership{Senior Member,~IEEE}

\thanks{Part of this work has been presented at ICIC 2024~\cite{wang2024improving}.}
\thanks{This work was supported in part by the National Natural Science Foundation of China with Grant 62301499, the Henan Association for Science and Technology with Grant 2025HYTP037, and the Open Foundation of Key Laboratory of Cyberspace Security, Ministry of Education of China with Grant KLCS20240211. \textit{(Corresponding author: Yan Cao.)}}
\thanks{Y. Huang, W. Wang, D. Zhang, J. Wang, C. Chen, and Y. Cao are with the School of Cyber Science and Engineering, Zhengzhou University, Zhengzhou 450001, China (e-mail: yonghuang@zzu.edu.cn, wwjj@gs.zzu.edu.cn, iedlzhang@zzu.edu.cn, junjiewang@gs.zzu.edu.cn, c1518763280@gmail.com, ieycao@zzu.edu.cn).}
\thanks{Wei Wang is with the School of Computer Science, Wuhan University, Wuhan 430072, China (e-mail: wangw@whu.edu.cn).}
\thanks{Copyright (c) 2025 IEEE. Personal use of this material is permitted. However, permission to use this material for any other purposes must be obtained from the IEEE by sending a request to pubs-permissions@ieee.org.}
}


\markboth{IEEE Internet of Things Journal, ~Vol.~X, No.~X, XXXXX~XXXX}%
{Shell \MakeLowercase{\textit{et al.}}: A Sample Article Using IEEEtran.cls for IEEE Journals}

\IEEEpubid{0000--0000/00\$00.00~\copyright~2025 IEEE}

\maketitle

\begin{abstract}
Radio frequency (RF) fingerprinting techniques provide a promising supplement to cryptography-based approaches but rely on dedicated equipment to capture in-phase and quadrature (IQ) samples, hindering their wide adoption.
Recent advances advocate easily obtainable channel state information (CSI) by commercial WiFi devices for lightweight RF fingerprinting, while falling short in addressing the challenges of coarse granularity of CSI measurements in an open-world setting.
In this paper, we propose CSI\textsuperscript{2}Q, a novel CSI fingerprinting system that achieves comparable performance to IQ-based approaches.
Instead of extracting fingerprints directly from raw CSI measurements, CSI\textsuperscript{2}Q first transforms frequency-domain CSI measurements into time-domain signals that share the same feature space with IQ samples. 
Then, we employ a deep auxiliary learning strategy to transfer useful knowledge from an IQ fingerprinting model to the CSI counterpart.
Finally, the trained CSI model is combined with an OpenMax function to estimate the likelihood of unknown ones.
We evaluate CSI\textsuperscript{2}Q on one synthetic CSI dataset involving 85 devices and two real CSI datasets, including 10 and 25 WiFi routers, respectively.
Our system achieves accuracy increases of at least 16\% on the synthetic CSI dataset, 20\% on the in-lab CSI dataset, and 17\% on the in-the-wild CSI dataset.
\end{abstract}


\section{Introduction}
\IEEEPARstart{N}{owadays}, WiFi has become one of the most essential communication technologies that connect various wireless devices, such as smartphones, tablets, and smart speakers.
Although WiFi has been indispensable in many aspects of our daily lives, it is susceptible to identity-based attacks like spoofing and impersonation.
The IEEE 802.11 protocol has provided cryptography-based authentication schemes, but their effectiveness has proven to be inadequate~\cite{36, 37}.
In recent years, radio frequency (RF) fingerprinting has been proposed as an important supplement to cryptography-based security technologies in wireless networks~\cite{44, 45, 4, 5, 13}.
It utilizes hardware imperfections, such as carrier frequency offsets in oscillators and nonlinearity in power amplifiers, to identify mobile devices~\cite{42}.
These hardware impairments are referred to as RF fingerprints, which are device-specific and difficult to forge~\cite{13}.
Additionally, compared to cryptography-based approaches, RF fingerprinting has lower requirements on computational complexity and communication overhead~\cite{6}, making it suitable for many miniaturized WiFi devices.

Currently, RF fingerprinting solutions rely on time-domain in-phase and quadrature (IQ) samples.
Handcrafted features are extracted from IQ signals as RF fingerprints for device identification~\cite{8, 13, 14, 15, 26, 27, 41}.
Additionally, raw or processed IQ signals are directly fed into deep neural networks for automatic RF fingerprint extraction~\cite{9, 10, 16, 18, 20, 28, 40, 43}.
While IQ-based solutions generally demonstrate high identification performance, their practical application in the real world is impeded by their requirement for costly and dedicated RF equipment like universal software radio peripherals (USRPs).

Growing attempts have been devoted to exploring RF fingerprinting based on channel state information (CSI) that is easily obtainable for commercial WiFi devices~\cite{29, 30, 31}.
A CSI measurement mainly describes how a WiFi signal propagates from a transmitter to a receiver, but it also carries RF fingerprints due to hardware imperfections.
Existing CSI fingerprinting schemes mainly focus on mitigating channel interference.
Some studies demonstrate that the channel-independent features, such as the carrier frequency offsets (CFOs)~\cite{21}, phase errors~\cite{22}, and power variances caused by the power amplifier~\cite{7}, exist in CSI measurements.
The literature~\cite{23, 24, 25} realizes CSI-based RF fingerprinting by eliminating channel interference.
However, besides channel interference, the task of WiFi CSI fingerprinting still faces two other challenges, and none of the existing works have addressed them.
\IEEEpubidadjcol
\begin{itemize}
    \item \textbf{Coarse Granularity.}
    According to the IEEE 802.11 protocol, a WiFi packet contains three parts, i.e., Preamble, SIGNAL, and DATA, in the time domain.
    Only IQ samples in the preamble are used for channel estimation.
    Even worse, one CSI measurement contains the channel responses of specified subcarriers.
    For example, for a 2.4~GHz WiFi with a 20~MHz bandwidth, the preamble signal consists of 320 IQ samples, but a CSI measurement has only 52 channel estimates of subcarriers.
    Thus, the coarse-grained feature representation renders it harder to extract subtle RF fingerprints from CSI measurements. 
    \item \textbf{Open-World Recognition.}
    Existing CSI fingerprinting schemes are typically designed for closed-world recognition~\cite{7, 21, 22, 23, 24, 25}, where the devices under test are also present during the training phase.
    However, the testing data coming from an unseen device can be misclassified as a known device, hampering the overall performance. 
    In practice, it is impossible to collect CSI measurements from all WiFi devices.
    Hence, a fingerprinting model that can recognize both known and unknown devices is desirable.
\end{itemize}

In this paper, we propose CSI\textsuperscript{2}Q, a novel CSI fingerprinting system that achieves comparable performance to IQ-based approaches. 
Besides channel interference, CSI\textsuperscript{2}Q tackles the other two challenges at the same time. 
The core idea of CSI\textsuperscript{2}Q is to transform frequency-domain CSI measurements into time-domain samples and transfer knowledge about RF fingerprints from an IQ-based model to its CSI-based counterpart via deep auxiliary learning.
To achieve this goal, we propose four effective components in our system.
\textit{First}, we perform phase unwrapping and correction to raw CSI measurements and propose a cyclic shift division scheme on all channel estimates in one CSI measurement for mitigating channel interference.
\textit{Second}, motivated by the relation between received preamble signals and true channel responses, we transform the processed CSI data into time-domain signals that share the same feature space with preamble IQ samples.
\textit{Third}, we introduce an auxiliary learning approach, where the distinct advantages of an IQ fingerprinting model in feature extraction are transferred to its CSI counterpart.
By doing so, the CSI fingerprinting model can capture fine-grained features that could be missed by it alone.
\textit{Last}, to achieve open-world recognition, we incorporate an OpenMax function at the output layer of the CSI fingerprinting model to calibrate the uncertainty of known devices and estimate the likelihood of unknown ones.

We build an auxiliary IQ dataset and a synthetic CSI dataset involving 85 wireless devices based on the public WiSig dataset~\cite{11}.
Moreover, we utilize a laptop running PicoScenes~\cite{31} to collect CSI measurements from commercial WiFi routers and obtain an in-lab CSI dataset consisting of 10 devices and an in-the-wild CSI dataset comprising 25 devices.  
We evaluate CSI\textsuperscript{2}Q on both the synthetic and real CSI datasets.
On the synthetic dataset, our system can improve the recognition accuracy from 79\% to 96\% in the closed-world scenario and from 72\% to 95\% in the open-world case.
On the in-lab CSI dataset, CSI\textsuperscript{2}Q boosts the accuracy from 66\% to 86\% in the closed-world scenario and from 54\% to 82\% in the open-world setting.
On the in-the-wild CSI dataset, CSI\textsuperscript{2}Q boosts the accuracy from 74\% to 90\% in the closed-world scenario and from 57\% to 82\% in the open-world setting.

The main contributions of this work are summarized as follows.
\begin{itemize}
    \item We show that CSI and IQ data are inherently correlated, and CSI fingerprinting can benefit from IQ samples for performance improvement.
    \item We propose CSI\textsuperscript{2}Q that addresses the challenges of coarse granularity and open-world recognition, achieving comparable device identification performance to IQ-based approaches.
    \item We evaluate CSI\textsuperscript{2}Q on both the synthetic and real CSI datasets.
    The experimental results show that our system can effectively improve the performance of CSI fingerprinting models.
\end{itemize}

\section{Threat Model and CSI Features}

\subsection{Threat Model}
We consider a common scenario where a WiFi access point (AP) is deployed to provide Internet access for a set of wireless devices, such as smart light bulbs, air conditioners, and so on.
Each device has a unique identity and is assigned a security key for successful association and authentication. 
In this scenario, we consider identity-based attacks, where a malicious device could impersonate another one by forging its identity, such as MAC and IP addresses, and stealing its security key.
After sneaking into the WiFi network, it will either inject fake data into a remote server or load the valuable files from it.

To defend against such attacks, we consider physical layer authentication based on radio frequency fingerprints.
First, we collect $N$ CSI measurements $\mathcal{H}=  \left\{H_n\right\}^{N}_{n=1} $ from $I$ registered devices $\mathcal{D} = \left\{D_i \right\}^{I}_{i=1} $.
Therein, $D_i\in  \left\{0,1\right\}^{I} $ is a one-hot vector indicating the device index.
Then, a classification model is trained based on the dataset $\mathcal{H}$.
Next, given a new CSI measurement $H$, the model recognizes which device the measurement comes from.
Moreover, we consider open-world recognition, where CSI measurements may belong to unknown devices that are unseen in $\mathcal{D}$ during testing.
Thus, the model classifies a CSI input into $I+1$ categories.

\subsection{CSI Features}
In this subsection, we introduce the basics of CSI features and conduct a preliminary study to show the feasibility and difficulty of WiFi CSI fingerprinting.

\textbf{CSI Basics.} 
Channel state information is utilized to characterize the propagation characteristics of 
wireless signals, reflecting their amplitude attenuation, phase variation, and time delay from the transmitter to the receiver.
Typically, the orthogonal frequency division multiplexing (OFDM) technique is adopted to divide the WiFi band into $K$ orthogonal subcarriers for high throughput.
In this condition, CSI is represented in a matrix form in the frequency domain.
Given the $k$-th subcarrier, let $x_k$ be the transmitted signal vector, and $y_k$ be the received signal vector. 
Moreover, $F(\cdot)$ indicates the impact of the transmitter's hardware imperfection, $\sigma$ represents the noise vector, and $h_{c_k}$ is the true channel response.
The relationship between $x_k$ and $y_k$ can be expressed as
\begin{equation}
  y_k = h_{c_k} \cdot F(x_k) + \sigma.
\end{equation}
The CSI of $k$-th subcarrier, denoted as $h_k$, is estimated by
\begin{equation}
  h_k = \frac{y_k}{x_k}
\end{equation}
Therefore, $h_k$ can be approximated as
\begin{align}
  h_k = {\frac{h_{c_k} \cdot F(x_k) + \sigma}{x_k}} \approx h_{c_k} \cdot {G(x_k)}, \label{H_k}
\end{align}
where $G(x_k) = \frac{F(x_k)}{x_k}$ includes the impact of the transmitter’s RF fingerprints. 
The above analysis shows that $h_k$ is an approximation of the true channel response $h_{c_k}$. 
Like IQ data, CSI data is also affected by the transmitter’s hardware imperfection, making it feasible to perform RF fingerprinting using WiFi CSI measurements.

\begin{figure}[t]
\centering
\captionsetup[subfigure]{font=scriptsize}
\subfloat[]{\includegraphics[width=0.48\linewidth]{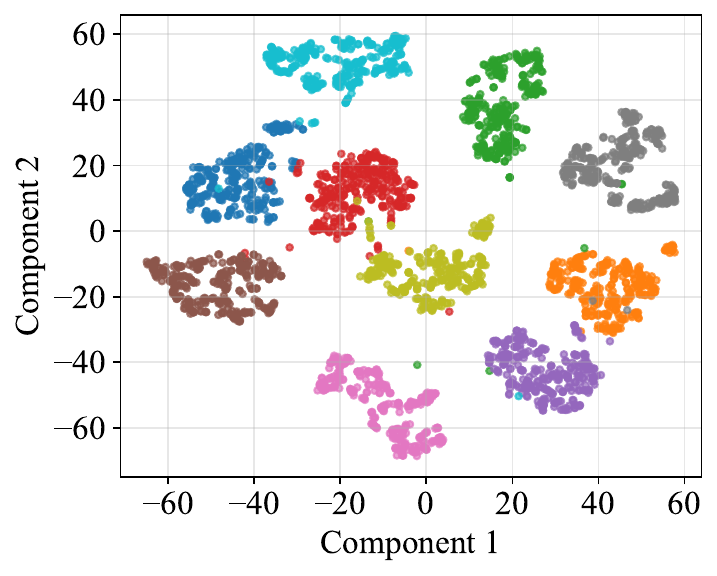}
\label{IQ_tsne_visualization}}
\hfil
\subfloat[]{\includegraphics[width=0.48\linewidth]{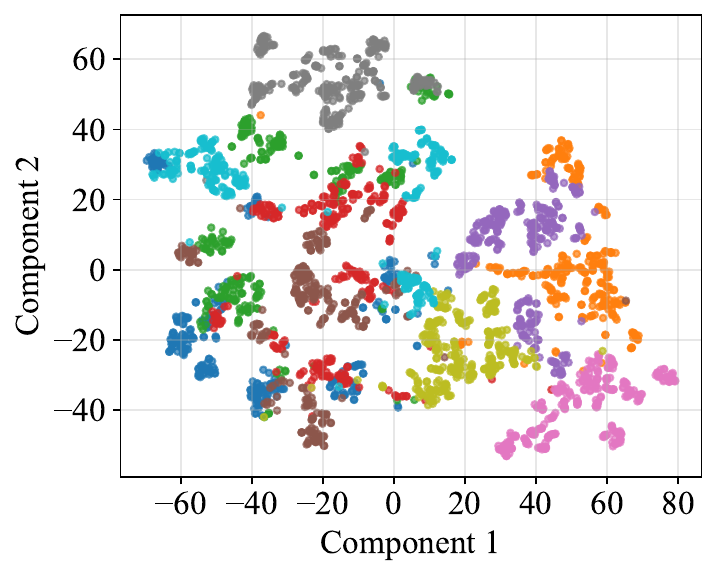}
\label{CSI_tsne_visualization}}
\caption{Visualization of high-dimensional features from IQ and CSI data. The same-color samples belong to one device. (a) IQ Data. (b) CSI Data.}
\label{fig:feature visualization}
\end{figure}

\textbf{Preliminary Study.}
We conduct some preliminary experiments to demonstrate the feasibility of CSI-based device classification.
To achieve this, we randomly take 3K IQ samples of ten devices from WiSig~\cite{11}, a recently released IQ dataset, and further calculate corresponding CSI measurements using the standard channel estimation process in MATLAB.
Then, we feed the two types of data into the same classification neural network, respectively, and exploit the t-distributed stochastic neighbor embedding (t-SNE) algorithm to visualize their high-dimensional features extracted by the network.
As shown in Fig.~\ref{fig:feature visualization}, the IQ data of the same device yield feature samples that are more tightly clustered.
Meanwhile, the boundaries between different clusters are more pronounced, making them easier to distinguish.
Although the feature samples from the CSI data also exhibit a clustering pattern, the CSI data of the same device are more dispersed and stay closer to that of other devices.
The above observation suggests that RF fingerprints are more difficult to extract from CSI features than IQ signals. 
In conclusion, CSI-based RF fingerprinting is feasible but more challenging than IQ-based approaches.

\section{System Design}

\subsection{System Overview}
CSI\textsuperscript{2}Q is a novel CSI-based RF fingerprinting system that achieves comparable performance to IQ-based approaches. 
The core idea of CSI\textsuperscript{2}Q is to transform frequency-domain CSI measurements into time-domain samples and transfer knowledge about RF fingerprints from an IQ-based model to its CSI-based counterpart via a deep auxiliary learning approach.
To realize this goal, the workflow of CSI\textsuperscript{2}Q consists of two phases, i.e., the training phase and the inference phase.
In the training phase, a CSI dataset from registered devices and an auxiliary IQ dataset are leveraged to train our fingerprinting model collaboratively.
In the inference phase, given a new CSI sample, the trained model recognizes which device the CSI sample comes from. 
\begin{figure}[t]
  \centering
  \includegraphics[width=\linewidth]{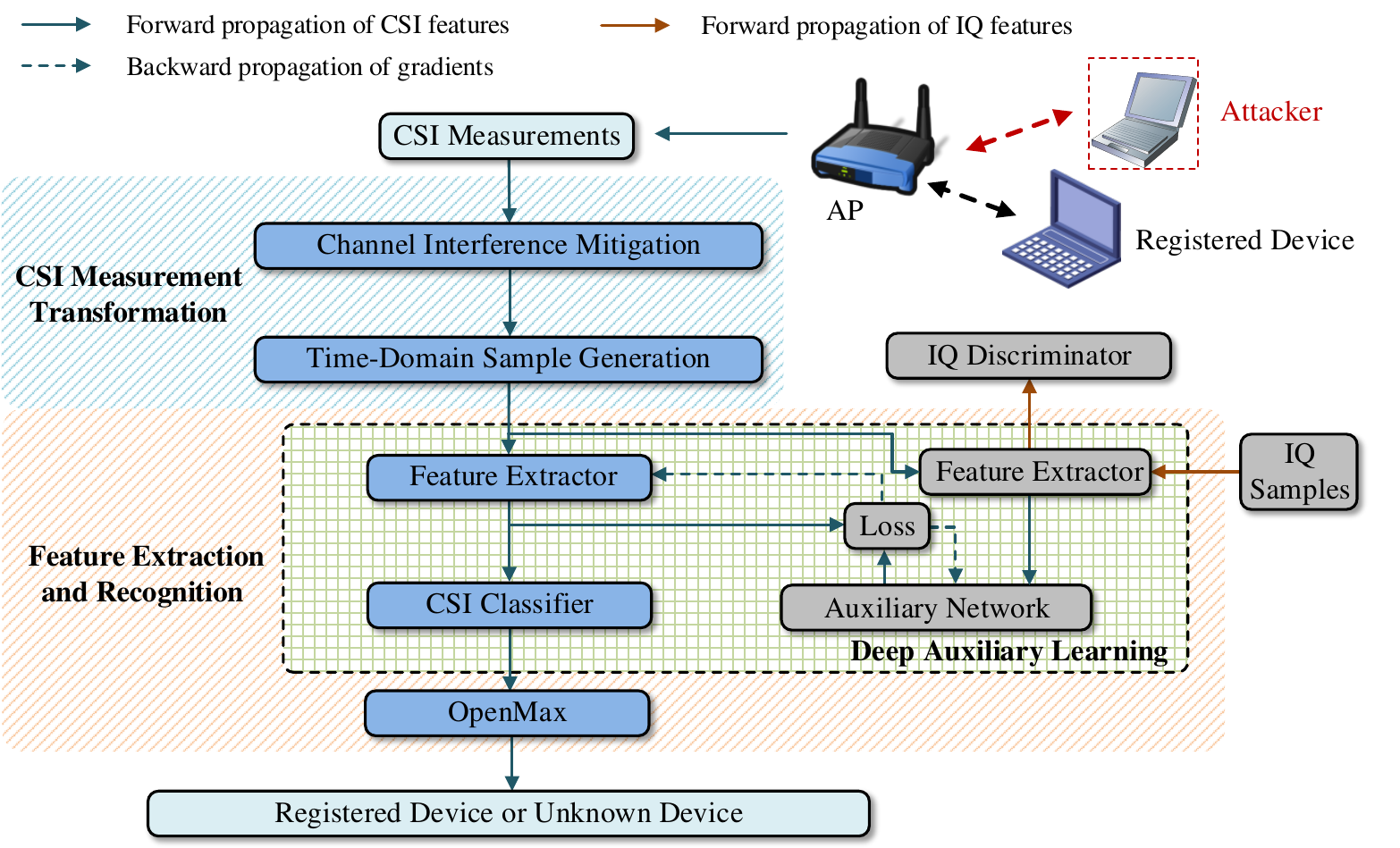}
  \caption{System overview of CSI\textsuperscript{2}Q. It consists of CSI measurement transformation and feature extraction and recognition. The dashed gray component exists only in the training phase.}
  \label{fig:system_design}
\end{figure}

As illustrated in Fig.~\ref{fig:system_design}, the proposed system consists of two main components -- CSI Measurement Transformation and Feature Extraction and Recognition.
\begin{itemize}
   \item \textbf{CSI Measurement Transformation.}
   First, we perform phase unwrapping and correct phase jitters on raw CSI measurements. 
   Then, we perform cyclic shift division on one CSI measurement for channel interference mitigation. 
   Next, we combine the processed CSI measurement with the standard short and long training sequences to generate a high-dimensional time-domain feature vector that shares the same feature space with the preamble IQ samples.
   \item \textbf{Feature Extraction and Recognition.}
   Based on the generated feature vector, we devise a deep auxiliary learning model with two feature extractors, an auxiliary network, a CSI classifier, and an IQ discriminator for effective feature extraction.
   In the training phase, a knowledge transfer approach is leveraged to improve the feature extraction ability of our CSI feature extractor and classifier.
   In the testing phase, we feed the classifier's output into an OpenMax function to recognize registered and unknown devices.
\end{itemize}

\subsection{CSI Measurement Transformation}
In this subsection, we transform highly dynamic and frequency-domain CSI measurements into channel-independent and time-domain features.
For ease of illustration, we take 2.4~GHz WiFi with a bandwidth of 20~MHz as an example to illustrate the process of feature transformation.
It is worth noting that the proposed CSI transformation method can be easily extended to other WiFi specifications.

\begin{figure}[t]
\centering
\captionsetup[subfigure]{font=scriptsize}
\subfloat[]{\includegraphics[width=0.48\linewidth]{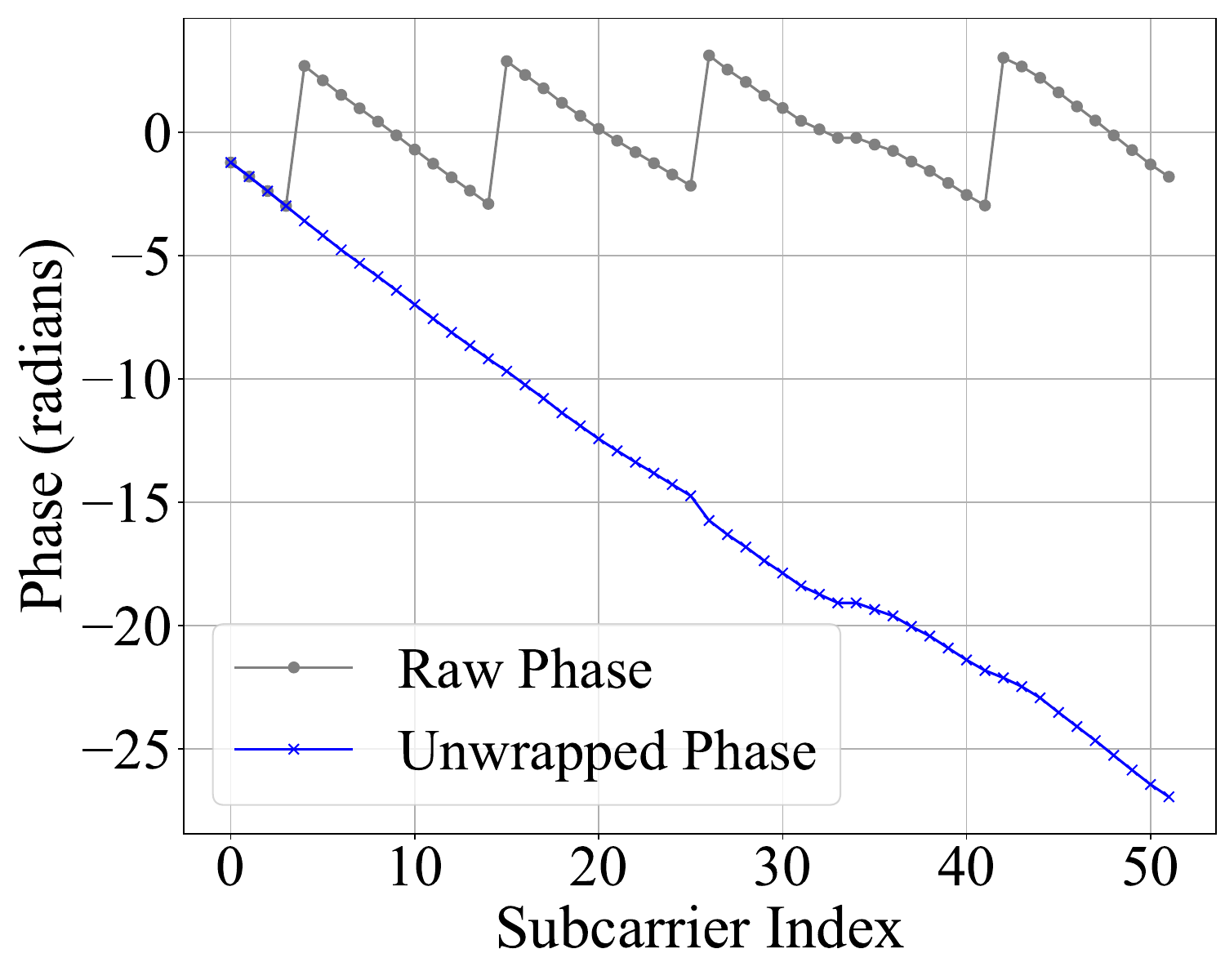}
\label{phase_unwarpping}}
\hfil
\subfloat[]{\includegraphics[width=0.48\linewidth]{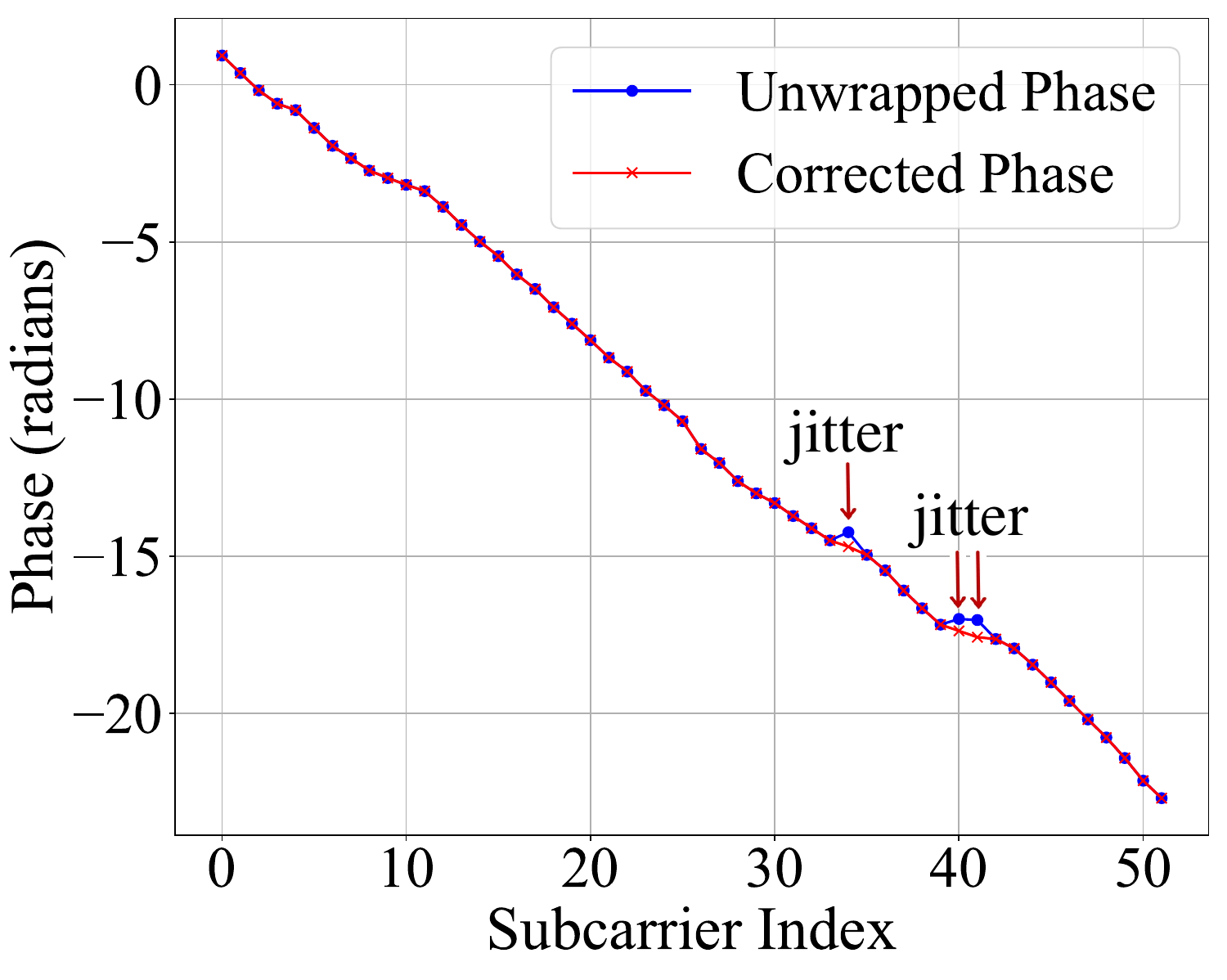}
\label{phase_correction}}
\caption{Phase unwrapping and phase correction. (a) Phase unwrapping. (b) Phase correction.}
\label{fig:csi_phase}
\end{figure}

\textbf{Channel Interference Mitigation.}
As aforementioned, a CSI measurement mainly reflects the characteristics of the propagating channel, and thus underlying transmitter RF fingerprints are highly disturbed. 
To deal with this issue, the first step of CSI\textsuperscript{2}Q is to mitigate channel interference in a CSI measurement. 
For a WiFi signal, let us denote $ K$ as the number of orthogonal subcarriers.
In this setting, a CSI measurement between one transmitting antenna and one receiving antenna consists of $K$ complex values, each of which corresponds to a frequency response of one subcarrier.
Formally, let $H \in \mathbb{C}^{K}$ be one CSI measurement, which can be denoted as
\begin{equation}
  H = [h_1,\ h_2,\ \dots,\ h_{K}].
\end{equation}
According to Eq.~\eqref{H_k}, one estimated channel response in $H $ can be expressed by
\begin{equation}
  h_k = h_{c_k} \cdot {G(x_k)},
  \label{H}
\end{equation}
where $k=1,\ 2,\ \dots,\ K$.
Generally, raw CSI phases are typically constrained within the range of $-\pi$ to $\pi$. 
As illustrated by the gray line in Fig.~\ref{fig:csi_phase}~(a), when the phase changes exceed this range, phase wrapping occurs.
To restore phase continuity, we detect abrupt phase jumps between -$\pi$ and $\pi$, calculate the required multiples of 2$\pi$ for adjustment, and apply them to the phase values.
The unwrapped phase curve is depicted as the blue line in Fig.~\ref{fig:csi_phase}~(a).

Next, we deal with phase jitters in CSI measurements. 
In an environment without significant disturbances, the subcarrier phase variations tend to exhibit a quasi-linear pattern within neighboring subcarriers.  
However, this quasi-linear relation can be disrupted by two categories of impact factors. 
The first category stems from hardware-induced characteristics, including oscillator phase noise and frequency instability~\cite{48}, residual synchronization errors caused by CFOs, sampling frequency offsets (SFOs), and sampling time offsets (STOs)~\cite{50}, 
as well as random phase offset among multiple RF chains~\cite{52}. 
For instance, the CFO contains a relatively stable component that can serve as a device-specific RF fingerprint. 
At the same time, its residual part after imperfect compensation leads to time-accumulated phase drifts that perturb the quasi-linear relation. 
The second category of influencing factors is environmental dynamics, where multipath effects make CSI phases highly sensitive to path delays; even small movements of surrounding people or objects may alter signal propagation paths, introducing Doppler-related instantaneous deviations~\cite{51,54}. 
Together, these factors result in irregular fluctuations in adjacent CSI phases, commonly referred to as phase jitters. 
Fig.~\ref{fig:csi_phase}~(b) presents a representative example of such jitters observed in one CSI measurement.
To address this issue, we identify the presence of jitters by analyzing phase gradients. 
When the phase gradient at a particular subcarrier changes in a direction opposite to that of its neighboring subcarriers, we consider that a phase jitter occurs.
If a CSI frame has fewer than five phase jitters, we smooth and correct disturbed phases using the values from adjacent subcarriers.
We discard this CSI frame if more than five subcarriers have phase jitters.
The red line in Fig.~\ref{fig:csi_phase}~(b) shows the corrected CSI phases, which are smoother than the unwrapped ones.

Then, we proceed to mitigate channel interference in CSI measurements.
Since the time duration of one WiFi packet falls within the channel coherence time, it can be assumed that the channel responses experienced by all subcarrier signals are highly correlated, especially for those of adjacent subcarriers.
If we divide one estimated channel response by another, the impact of channel characteristics is likely to be alleviated.
Thus, we use a cyclic shift division scheme for channel interference mitigation.
In this scheme, when $k= 2,\ \dots,\ K$, we divide $h_k$ by $h_{k-1}$ as
\begin{align}
  \widetilde{h_k} = \frac{h_k}{h_{k-1}} = \frac{{h_{c_k}} \cdot {G(x_k)}}{{h_{c_{k-1}}} \cdot {G(x_{k-1})} } \approx \frac{G(x_k)}{G(x_{k-1})}.
\end{align}
When $k=1$, we have
\begin{equation}
    \widetilde{h_1} = \frac{h_1}{h_{2}} \approx \frac{G(x_1)}{G(x_{2})}.
\end{equation}
From the above two equations, we can observe that the impact of channel characteristics could be removed, and the transmitter's RF fingerprints remain in $\widetilde{h_k}$.
After cyclic shift division on the CSI measurement $H$, a processed CSI measurement $\widetilde{H} \in \mathbb{C}^{K} $ can be obtained as
\begin{equation}
  \widetilde{H}=[\widetilde{h}_1,\ \widetilde{h}_2,\ \dots,\ \widetilde{h}_{K}]. 
  \label{H_1}
\end{equation}

\begin{figure}[t]
  \centering
  \includegraphics[width=\linewidth]{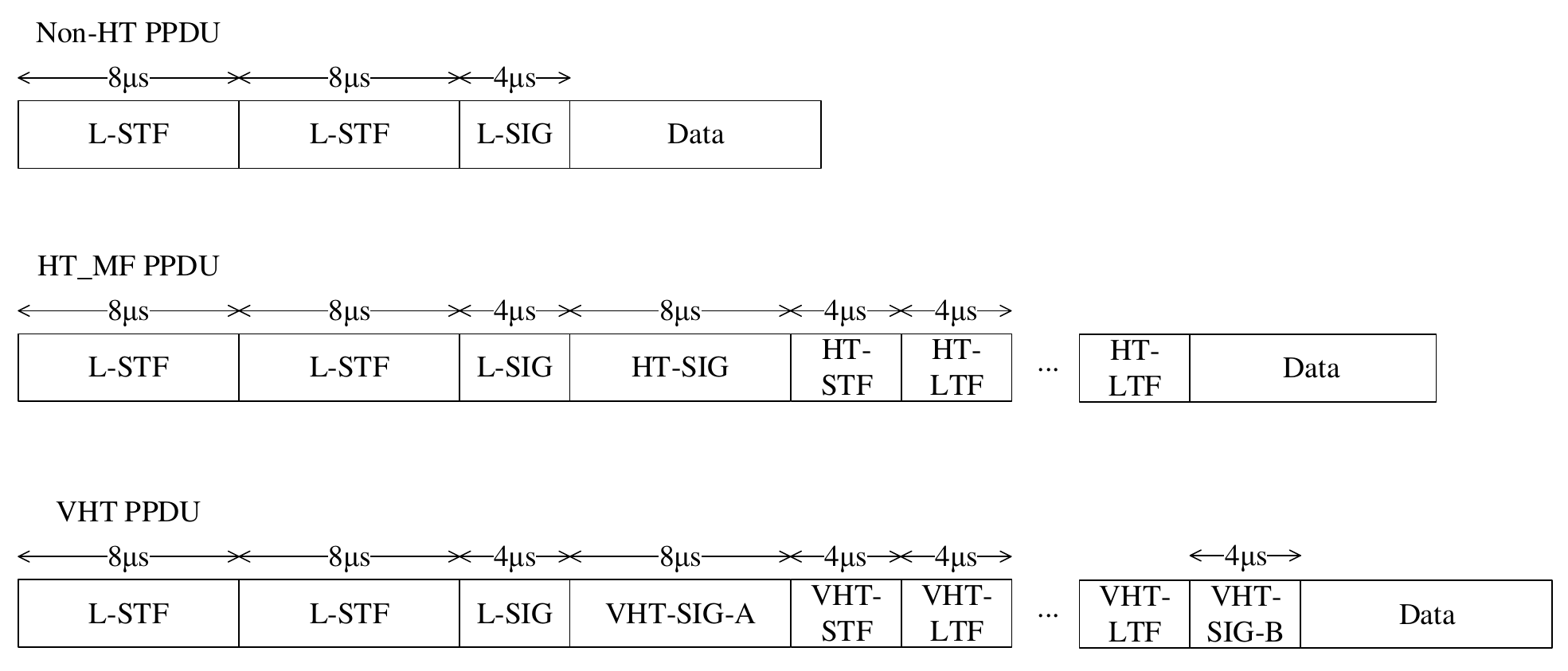}
  \caption{PPDU data frame structure with different formats. }
  \label{fig:PPDU_frame}
\end{figure}

\textbf{Time-Domain Sample Generation.} 
After channel interference mitigation, we proceed to transform a processed CSI measurement into time-domain features like IQ data.  
According to the IEEE 802.11 protocol, the physical layer convergence procedure (PLCP) preamble field in the physical layer protocol data unit (PPDU) frame is used for channel estimation.
As shown in the Fig.~\ref{fig:PPDU_frame}, although different versions of PPDU frames are different in physical layer implementation, they contain legacy short training field (L-STF) and legacy long training field (L-LTF) with fixed structures.
This design ensures that different protocol devices can realize basic communication functions by retaining training sequences with a fixed duration.
In addition, with the evolution of the protocol, the bandwidth of WiFi communication is also expanding, but the L-STF and L-LTF parts of its physical layer preamble still maintain the infrastructure consistent with the bandwidth of 20 MHz. 
Taking 40 MHz bandwidth as an example, its preamble consists of two parallel 20 MHz subchannels, each of which contains independent L-STF and L-LTF structures.
Similarly, 80 MHz bandwidth is achieved by splicing four 20 MHz subchannels.
The unification of this technical route provides a basis for cross-protocol version analysis of CSI fingerprint identification.
It is worth noting that the IEEE 802.11 protocol stipulates that under OFDM modulation, L-LTF sequences are distributed on 52 effective subcarriers out of 64 subcarriers, and the remaining 12 subcarriers are zeroed due to band protection requirements.
CSI measurements estimated by L-LTF generally only contain amplitude and phase information of 52 subcarriers.
Therefore, when the preamble signal is sampled at a rate of 20 Msps, 320 time-domain IQ sampling points can be obtained, which can be further used to generate CSI measurement values with 52 complex values.
Compared with the corresponding IQ data, the granularity of CSI measurements is coarser, which makes it more difficult to extract potential device fingerprints.
Based on the commonness between preambles of WiFi signals with different protocol versions and bandwidth settings, the processed CSI data is used to generate time domain feature vectors with the same length as the preamble IQ signal, so as to solve the problem of coarse granularity of CSI measurements.

Firstly, we model the signal containing the same L-STF and L-LTF preamble as
\begin{equation}
  s(t) = s_S(t)+s_L(t-T).
\end{equation}
where $s_S(t)$ is the short training signal, $s_L(t-T)$ is the long training signal, and $T$ is the short training signal duration.
Let $\Omega_S$ be the set of subcarrier indices for the L-STF, $\Delta_f$ be the subcarrier spacing, $S$ be the short training symbol, $s_k$ be the $k$-th element of $S$, $H=[h_1,h_2,...,h_K]$ be the channel impulse response, and $\omega_T(t)$ represent a window function extending over multiple periods of fast Fourier transform (FFT).
Then, the L-STF can be expressed as
\begin{equation}
  s_S(t) = w_{T}(t) \sum_{k \in \Omega_S} s_k h_k \exp(j2 \pi k {\Delta}_f t).
  \label{st}
\end{equation}
Similarly, let $\Omega_L$ be the set of subcarrier indices for the L-LTF, $L$ be the long training symbol, $l_k$ be the $k$-th element of $L$, $T_{GI2}$ be the training symbol guard interval duration, and then the L-LTF can be expressed as
\begin{equation}
  s_L(t) = w_{T}(t) \sum_{k \in \Omega_L} l_k h_k \exp(j2 \pi k {\Delta}_f (t-T_{GI2})).
  \label{lt}
\end{equation}
For each 20 MHz bandwidth subchannel, the short training symbol $S \in \mathbb{C}^{52}$ consists of 12 effective subcarriers.
In the frequency domain, it can be represented as
\begin{equation}
  \begin{aligned}
  S =& {\sqrt{(13/6)}} \times \left\{0,\ 0,\ 1+j,\ 0,\ 0,\ 0,\ -1-j,\ 0,\ 0,\ 0,\right.\\ 
  &\left. 1+j,\ 0,\ 0,\ 0,\ -1-j,\ 0,\ 0,\ 0,\ -1-j,\ 0,\ 0,\ 0,\right.\\ 
  &\left. 1+j,\ 0,\ 0,\ 0,\ 0,\ 0,\ 0,\ -1-j,\ 0,\ 0,\ 0,\ -1-j,\right.\\ 
  &\left. 0,\ 0,\ 0,\ 1+j,\ 0,\ 0,\ 0,\ 1+j,\ 0,\ 0,\ 0,\ 1+j,\ 0,\right.\\ 
  &\left. 0,\ 0,\ 1+j,\ 0,\ 0\right\},
    \end{aligned}
    \label{S}
\end{equation}
where $\sqrt{(13/6)}$ is a power normalization factor.
The window function $\omega_T(t)$ is defined as
\begin{equation}
  w_{T}(t) = 
  \begin{cases}
  \sin^2(\frac{\pi}{2}(0.5+{\frac{t}{T_{TR}}})), &  - {\frac{T_{TR}}{2}}<t<\frac{T_{TR}}{2};\\
  1, &  {\frac{T_{TR}}{2}}\leq t<T-\frac{T_{TR}}{2};\\
  \sin^2(\frac{\pi}{2}(0.5-{\frac{t-T}{T_{TR}}})), &  {T-\frac{T_{TR}}{2}}\leq t< T+\frac{T_{TR}}{2},
  \end{cases}
  \label{w_T}
\end{equation}
where $T_{TR}$, the transition time, is about 100~ns and $T = 10 \times 0.8 \mu s = 8 \mu s$.
Then we take Eq.~(\ref{S}), Eq.~(\ref{w_T}), and Eq.~(\ref{H_1}) (processed CSI data $ \widetilde{H}$) into Eq.~(\ref{st}), the L-STF signal with transmitter's RF fingerprints can be obtained as
\begin{equation}
  x_S(t) = w_{T}(t) \sum_{k=-26}^{26} s_k \widetilde{h}_k \exp(j2 \pi k {\Delta}_f t),
\end{equation}
where ${\Delta}_f=312.5$~kHz is the frequency spacing between adjacent subcarriers.

The long training symbol $L \in \mathbb{C}^{52}$ uses all 52 subcarriers, and it can be expressed as
\begin{equation}
  \begin{aligned}
  L=& \left\{1,\ 1,\ -1,\ -1,\ 1,\ 1,\ -1,\ 1,\ -1,\ 1,\ 1,\ 1,\ 1,\ 1,\ 1,\right.\\ 
  &\left. -1,\ -1,\ 1,\ 1,\ -1,\ 1,\ -1,\ 1,\ 1,\ 1,\ 1,\ 1,\ -1,\ -1,\right.\\ 
  &\left. 1,\ 1,\ -1,\ 1,\ -1,\ 1,\ -1,\ -1,\ -1,\ -1,\ -1,\ 1,\ 1,\right.\\ 
  &\left. -1,\ -1,\ 1,\ -1,\ 1,\ -1,\ 1,\ 1,\ 1,\ 1\right\}.
    \end{aligned}
\end{equation}
Similarly, the L-LTF signal with RF fingerprint indices can be obtained by
\begin{equation}
  x_L(t) = w_{T}(t) \sum_{k=-26}^{26} l_k \widetilde{h}_k \exp(j2 \pi k {\Delta}_f (t-T_{GI2})),
\end{equation}
where $T_{GI2}=1.6~\mu s $ is the guard interval for the training symbol.

\begin{figure}[t]
\centering
\captionsetup[subfigure]{font=scriptsize}
\subfloat[]{\includegraphics[width=0.48\linewidth]{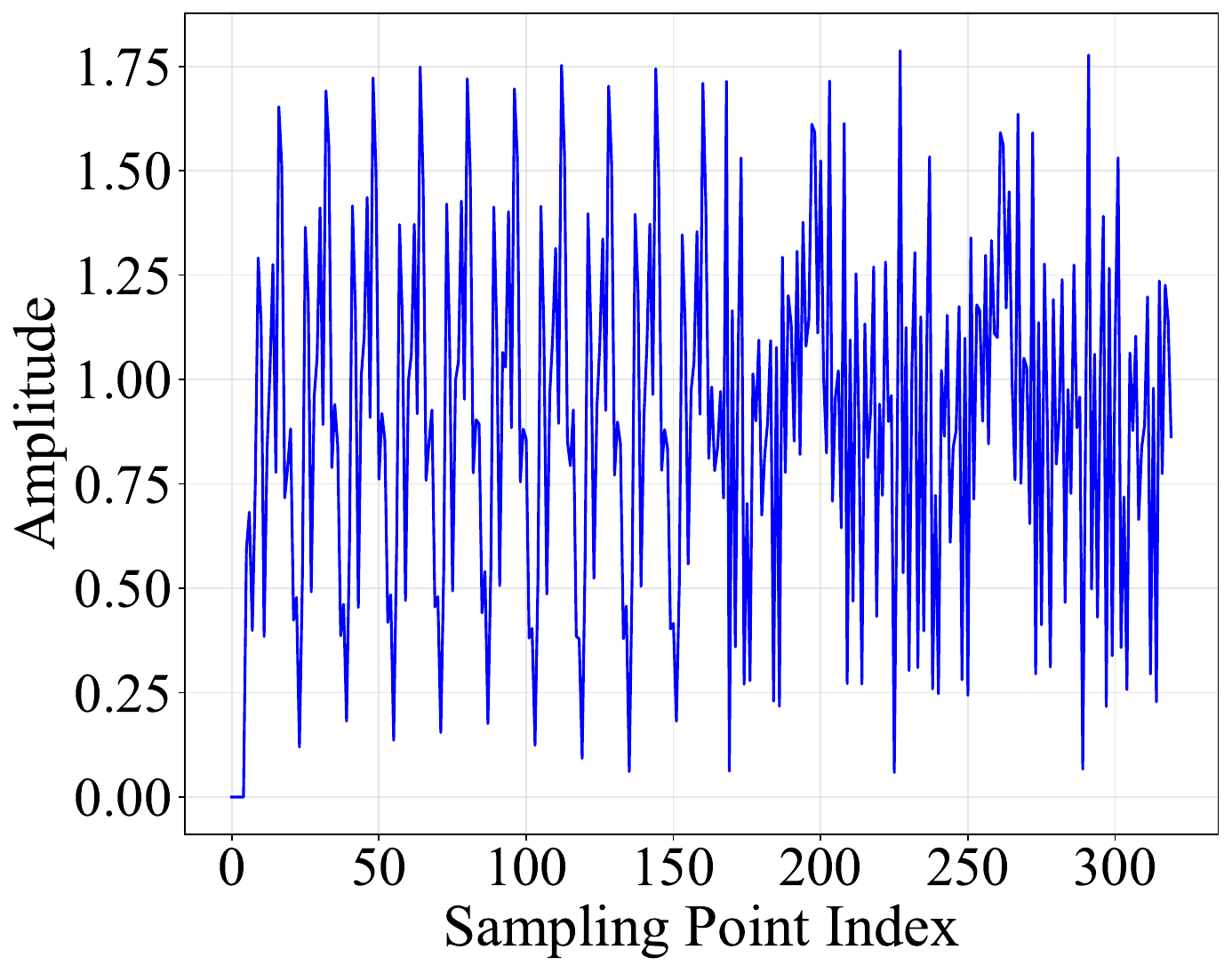}
\label{IQ_sample_amplitude_visualization}}
\hfil
\subfloat[]{\includegraphics[width=0.48\linewidth]{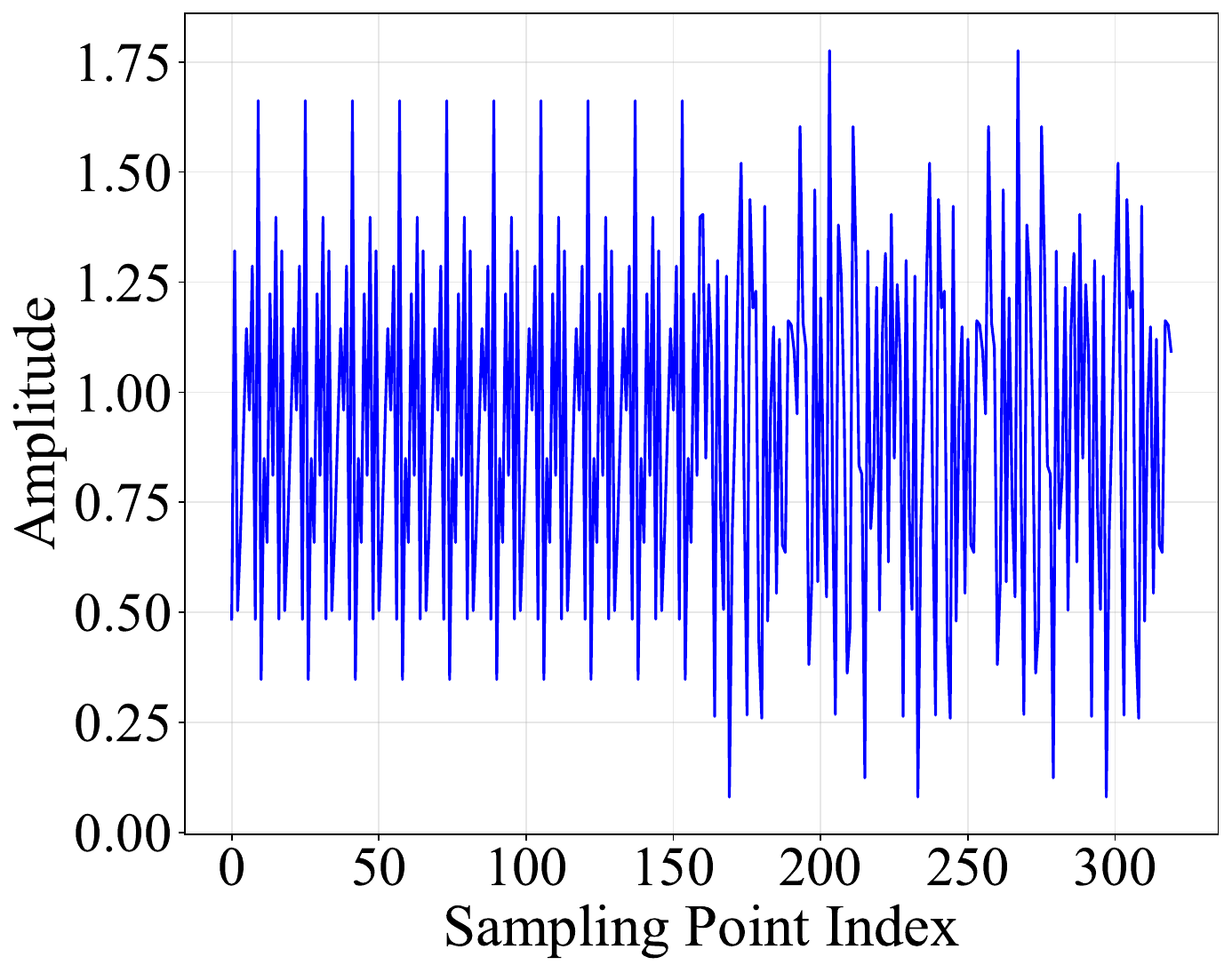}
\label{sltf_eq_sample_amplitude_visualization}}
\caption{Real preamble IQ signal and generated time-domain CSI feature vector. (a) Preamble IQ signal. (b) Time-domain CSI feature vector.}
\label{fig:time_domain_sample}
\end{figure}

Toward this end, we add $x_S(t)$ and $x_L(t)$ together and have the generated time-domain signal as
\begin{equation}
  x(t) = x_S(t)+x_L(t-T).
\end{equation}
Finally, by sampling it at a rate of 20~Msps, we obtain a time-domain feature vector $U\in \mathbb{C}^{320}$ as
\begin{equation}
      U = [u_1,\ u_2,\ \dots,\ u_{320}].
\end{equation}
In this way, we convert a CSI measurement of 52 subcarriers into a high-dimensional feature vector, which has the same length as a time-domain preamble signal and carries RF hardware imperfection characteristics at the same time.
We depict an IQ sample and the corresponding CSI feature vector in Fig.~\ref{fig:time_domain_sample}.
As the figure shows, two types of signals share the same feature space and have a similar changing pattern, which lays the foundation for transferring useful knowledge from IQ samples to CSI features.

\subsection{Feature Extraction and Recognition}
In this subsection, we perform auxiliary learning-based feature extraction and open-world recognition based on the generated time domain feature vectors.

\begin{figure}[t]
  \centering
  \includegraphics[width=\linewidth]{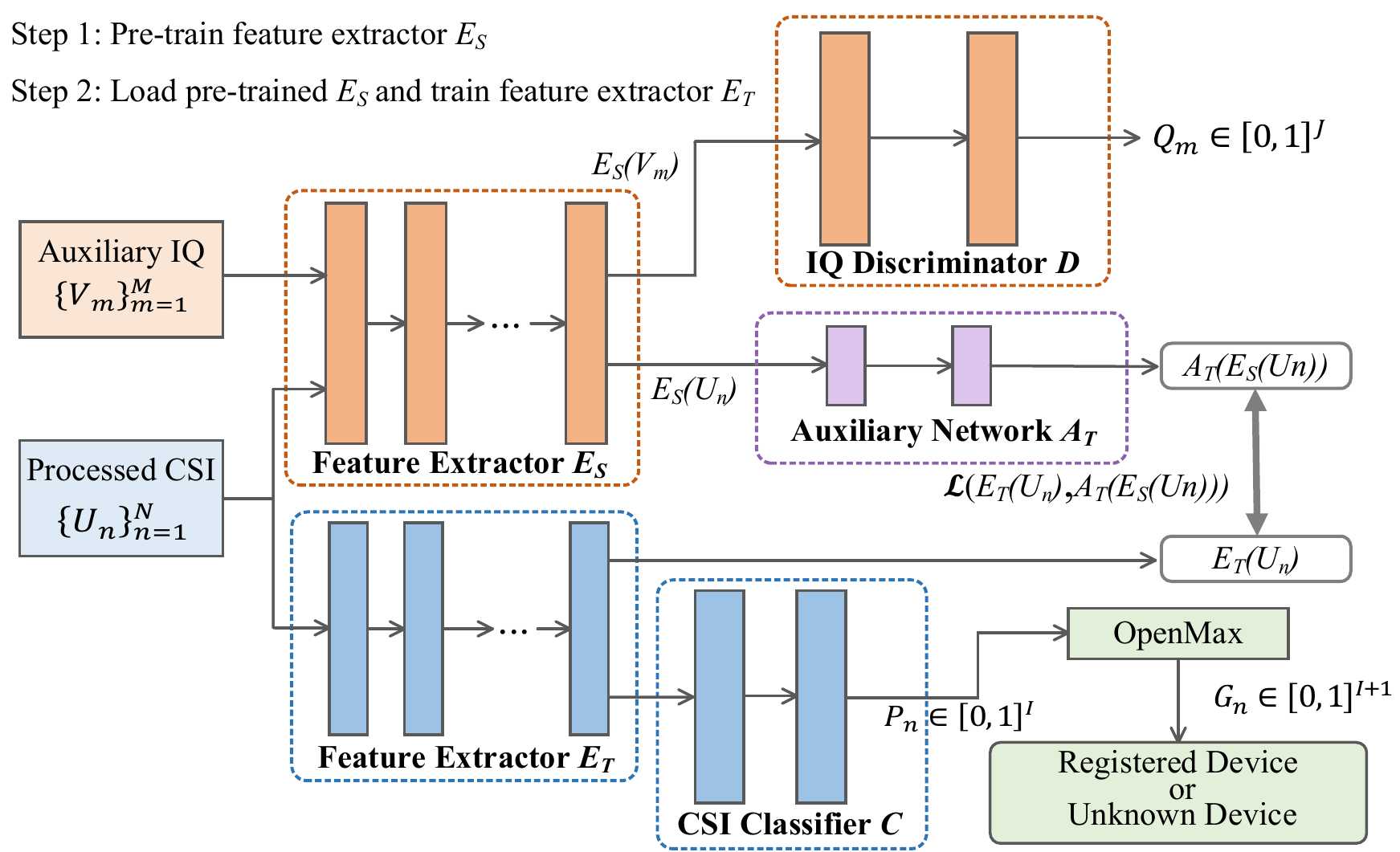}
  \caption{Workflow of the auxiliary knowledge transfer model.}
  \label{fig:model_architecture}
\end{figure}

\textbf{Auxiliary Learning with IQ Samples.}
Although CSI measurements are transformed into higher-dimensional feature vectors, they are not authentic IQ signals and need further feature extraction.
To achieve this, we exploit an auxiliary learning approach for CSI-based RF fingerprinting.
Our goal is to transfer the feature representation learned from the IQ samples to the CSI feature vector that shares the same feature space.
Specifically, after the CSI measurement transformation, we consider the task of fingerprinting registered devices as a multiclass classification problem $(\mathcal{U},\mathcal{D})$.
Therein, $\mathcal{U} = \left\{U_n\right\}^{N}_{n=1} \subseteq \mathbb{C}^{320} $ is the dataset of CSI feature vectors and can be extracted from the collected CSI measurements $\mathcal{H} =  \left\{H_n\right\}^{N}_{n=1} $.
In addition, we define an auxiliary IQ dataset $\mathcal{V} = \left\{V_m\right\}^{M}_{m=1} \subseteq \mathbb{C}^{320} $ that contains raw IQ samples from $J$ wireless devices $\mathcal{D}^v$.
Note that we do not assume any relation between two device sets $\mathcal{D}$ and $\mathcal{D}^v$.
For instance, $\mathcal{D}$ could be a subset of  $\mathcal{D}^v$, i.e., $\mathcal{D} \subseteq \mathcal{D}^v$, or they do not have common elements, i.e., $\mathcal{D} \cap \mathcal{D}^v = \varnothing$. 

For effective device classification, we propose an auxiliary knowledge transfer model, as shown in Fig.~\ref{fig:model_architecture}.
In particular, the proposed model consists of five components, including two feature extractors $E_S$ and $E_T$, an auxiliary network $A_T$, a CSI classifier $C$, and an IQ discriminator $D$. 
The basic idea is that since CSI measurements and IQ samples are inherently correlated, the shared feature extractor $E_S$ and auxiliary network $A_T$ can help transfer knowledge from IQ samples to the CSI feature extractor $E_T$ and classifier $C$ when processing time-domain CSI feature vectors. 
The shared feature extractor $E_S$ can provide additional insights and allow the feature extractor to capture fine-grained details that may be missed by $E_S$ alone.
There are two major steps in our auxiliary learning approach.

In the first step, we train the feature extractor $E_S$ along with the IQ discriminator $D$ on the source domain $\mathcal{V} $, making $E_S$ obtain the feature extraction ability of IQ samples. 
To achieve this, IQ samples $\left\{V_m\right\}^{M}_{m=1}$ are fed into $E_S$, and $ D $ performs IQ fingerprinting by yielding a $J$-dimensional probability vector $Q_m \in [0,1]^{J}$. 
We adopt the cross-entropy loss to train $E_S$ and $D$.
Generally, the cross-entropy calculates the average difference between predicted probabilities and ground-truth labels and has the following advantages.
First, it penalizes incorrect predictions more heavily, forcing the model to focus on learning correct classes.
Second, it provides a smooth and continuous optimization landscape, facilitating faster convergence during training.
Hence, we learn the parameters of $E_S$ and $D$ by minimizing their cross-entropy loss on the dataset $\mathcal{V} $ as
\begin{equation}
    {\min} \: \mathcal{L}_{source}(E_S,D),
\end{equation}
where
\begin{equation}
    \mathcal{L}_{source}(E_S,D) = -{\frac{1}{M}} \sum_{m=1}^{M} \sum_{j=1}^{J} D^j_m \log(Q^j_m).
\end{equation}
Therein, $D^j_m$ denotes the binary indicator (0 or 1) of whether the $m$-th IQ data belongs to the $j$-th device, and $Q^j_m$ represents the predicted probability of the $m$-th data belonging to the $j$-th device.

In the second step, we transfer the useful knowledge about IQ feature extraction to the target feature extractor $E_T$ and boost the classification performance of the CSI classifier $C$.
We first load the trained $E_S$ and put a target domain data $U_n$ into $E_S$ and $E_T$ respectively, thus obtaining the feature vectors $E_S(U_n)$ and $E_T(U_n)$.

On the one hand, the classifier $C$ takes a learned feature representation $E_T(U_n)$ as input and outputs an $I$-dimensional probability vector $P_n \in [0,1]^{I}$ in terms of all registered devices.
Similarly, we also choose the cross-entropy loss function for the classifier $C$.
The loss function of the target task $\mathcal{L}_{target}(E_T, C)$ is given as follows
\begin{equation}
  \mathcal{L}_{target}(E_T, C) = -{\frac{1}{N}} \sum_{n=1}^{N} \sum_{i=1}^{I} D^i_n \log(P^i_n),
\end{equation}
where $D^i_n$ denotes the binary indicator (0 or 1) of whether the $n$-th CSI feature sample belongs to the $i$-th device, and $P^i_n$ represents the predicted probability of the $n$-th sample belonging to the $i$-th device.

On the other hand, we train the auxiliary network $A_T$ to align features from the two feature extractors $E_T$ and $E_S$.
To achieve this goal, the auxiliary network $A_T$ takes the feature vector $E_S(U_n)$ as input and outputs an intermediate feature vector $A_T(E_S(U_n))$ that has the same dimension as the feature vector $E_T(U_n)$. 
For effective training, we aim to minimize the distance between $E_S(U_n)$ and $A_T(E_S(U_n))$.
It is worth noting that we introduce the auxiliary network $A_T$ into $E_S$, because directly aligning the outputs of $E_S$ and $E_T$ is ineffective due to data distribution discrepancy between the source data domain $\mathcal{V}$ and target data domain $\mathcal{U}$~\cite{39}.
Moreover, the re-training process could weaken the feature extraction ability of $E_S$.
Thus, during this auxiliary training, we freeze the parameters of $E_S$ and gradually transfer its knowledge to $E_T$ by training the parameters of $A_T$. 
Specifically, their features are aligned by minimizing the mean squared error (MSE) between $A_T(E_S(U_n))$ and $E_T(U_n)$, which can be expressed as
\begin{equation}
  \mathcal{L}_{auxi} (E_T, A_T) = -{\frac{1}{N}} \sum_{n=1}^{N} (E_T(U_n)-A_T(E_S(U_n)))^2.
\end{equation}
Finally, we implement the auxiliary learning approach by solving the following minimization problem as
\begin{equation}
    {\min} \: \mathcal{L}_{target} (E_T, C) + \lambda \mathcal{L}_{auxi} (E_T, A_T),  
\end{equation}
where $\lambda>0$ is a hyperparameter, weighting the importance of two losses.

\textbf{Open-World Recognition.}
In practice, any WiFi device could trigger an authentication request to the AP.
Thus, it is infeasible to collect CSI measurements from all possible devices.
To deal with this issue, we employ an OpenMax function~\cite{32} on the output of the CSI classifier $C$ to extend its $I$-dimensional probability vector $ P_n $ into a calibrated one $ G_n $ with $I+1$ dimensions to facilitate open-world recognition.
Mathematically, given the time-domain feature vector $U_n$, this function converts the output of the main task model $ P_n $ into $ G_n $ as
\begin{align}\label{eq:calibrated_vector}
    G_n  = [p_1 c_1, \ p_2 c_2, \ \cdots, \ p_I c_I, \ g].
\end{align}
Therein, $c_i \in [0,1]$ is a confidence weight, indicating the likelihood that $U_n$ belongs to the $i$-th device.
$g$ indicates the probability of unknown devices and is defined as the sum of increased uncertainty about known devices, which is given by
\begin{align} \label{eq:q-value}
   g = \sum_{i=1}^{I} p_i (1-c_i).
\end{align}
Based on Eq.~\eqref{eq:calibrated_vector} and Eq.~\eqref{eq:q-value}, $G_n$ is also a probability vector, and the sum of all its elements is equal to 1. 
To evaluate $c_i$, the activation vector of $U_n$, i.e., the output of the last fully connected layer in the CSI classifier, is compared with that of correctly classified training samples of the $i$-th category using the Weibull distribution method.
Generally, the closer they are, the bigger the $c_i$ is.
Eq.~\eqref{eq:calibrated_vector} suggests that the OpenMax function can model the feature space of registered devices and enables the model to discriminate unknown ones, thus improving its performance in open-world recognition.
Based on the calibrated probability vector $ G_n $, the OpenMax function selects the most likely device $i_n$ as
\begin{align}
    i_n = \left\lbrace 
    \begin{array}{lcl}
    I+1 \: , & \text{if} \: \: {g > \delta} ; \\
    \arg \underset{{i=1:I}}{\max} \: G_n, \:  & \text{else} .
\end{array}
\right.
\end{align}
Therein, $\delta$ is a threshold. 
When $g >\delta$, the model considers that the sample $U_n$ is from an unknown device. Otherwise, it can be considered from the registered device with the highest probability. 
We summarize the above processes in Algorithm~\ref{alg:training_with_openmax}.

To this end, given a new CSI measurement, the trained feature extractor and CSI classifier can effectively recognize both registered and unknown devices.
It is worth noting that the parameters of the deep auxiliary learning model are not specified.
Hence, CSI\textsuperscript{2}Q can be considered as a general performance improvement framework that generalizes to other neural network specifications.

\begin{algorithm}[t]
\caption{Deep Auxiliary Learning with OpenMax}\label{alg:training_with_openmax}
\begin{algorithmic}
\STATE \textbf{Input:} Time-domain CSI feature vectors and labels $(U_n, Y_{U_n})$, auxiliary IQ samples and labels $(V_m, Y_{V_m})$, hyperparameters $\eta_{E_S}$, $\eta_{E_T}$, $\eta_{A_T}$, $epoch_S$, $epoch_T$, threshold $\delta$
\STATE

\STATE \textbf{Step 1: Pretrain Source Model $E_S$ and $D$}
\STATE Initialize optimizer for $E_S$ with $\eta_{E_S}$
\FOR{$\text{epoch} \gets 1$ \textbf{to} $epoch_S$}
    \FOR{each $(V_m^{\text{batch}}, Y_{V_m}^{\text{batch}}) \in (V_m, Y_{V_m})$}
        \STATE $o_S \gets D(E_S(V_m^{\text{batch}}))$
        \STATE $\mathcal{L}_{source} \gets \text{CrossEntropy}(o_S, Y_{V_m}^{\text{batch}})$
        \STATE Update $E_S$ and $D$ with $\nabla \mathcal{L}_{source}$
    \ENDFOR
\ENDFOR

\STATE
\STATE \textbf{Step 2: Train Target Model $E_T$ and $C$ with Auxiliary Network $A_T$}
\STATE Initialize optimizers for $E_T$ and $A_T$ with $\eta_{E_T}$, $\eta_{A_T}$
\FOR{$\text{epoch} \gets 1$ \textbf{to} $epoch_T$}
    \FOR{each $(U_n^{\text{batch}}, Y_{U_n}^{\text{batch}}) \in (U_n, Y_{U_n})$}
        \STATE $e_T \gets E_T(U_n^{\text{batch}})$
        \STATE $o_T \gets D(e_T)$
        \STATE $\mathcal{L}_{\text{target}} \gets \text{CrossEntropy}(o_T, Y_{U_n}^{\text{batch}})$
        \STATE $e_S \gets E_S(U_n^{\text{batch}})$
        \STATE $e_A \gets A_T(e_S)$
        \STATE $\mathcal{L}_{auxi} \gets \text{MSE}(e_A, e_T)$
        \STATE Update $A_T$ with $\nabla \mathcal{L}_{auxi}$
        \STATE Update $E_T$ and $C$ with $\nabla \mathcal{L}_{\text{target}}+\lambda \nabla \mathcal{L}_{auxi}$
    \ENDFOR
\ENDFOR

\STATE
\STATE \textbf{Step 3: Apply OpenMax for Open-World Recognition}
\STATE $\alpha, \beta \gets \text{WeibullFit}(o_T)$
\FOR{each test sample $U_n^{\text{test}}$}
    \STATE ${o}_T^{\text{test}} \gets D(E_T(U_n^{\text{test}}))$
    \STATE $P_n =[p_i, p_2, ..., p_I] \gets \text{Softmax}({o}_T^{\text{test}})$
    \FOR{$i \gets 1$ \textbf{to} $I$}
        \STATE $c_i \gets 1- \text{WeibullCDF}(\alpha, \beta, p_i)$
    \ENDFOR
    \STATE $g \gets \sum_{i=1}^{I}{(pi * (1 - ci))}$
    \STATE $G_n \gets [p_1*c_1, p_2*c_2, ..., p_I*c_I, g]$
    \STATE Compare $g$ with $\delta$ to identify devices
\ENDFOR

\STATE
\STATE \textbf{Output:} Device ID
\end{algorithmic}
\end{algorithm}

\section{Experimental Evaluation}
For simplicity, we use CIM, TDSG, and ALIQ to represent channel interference mitigation, time-domain sample generation, and auxiliary learning with IQ, respectively.

\begin{figure}[t]
  \centering
  \includegraphics[width=\linewidth]{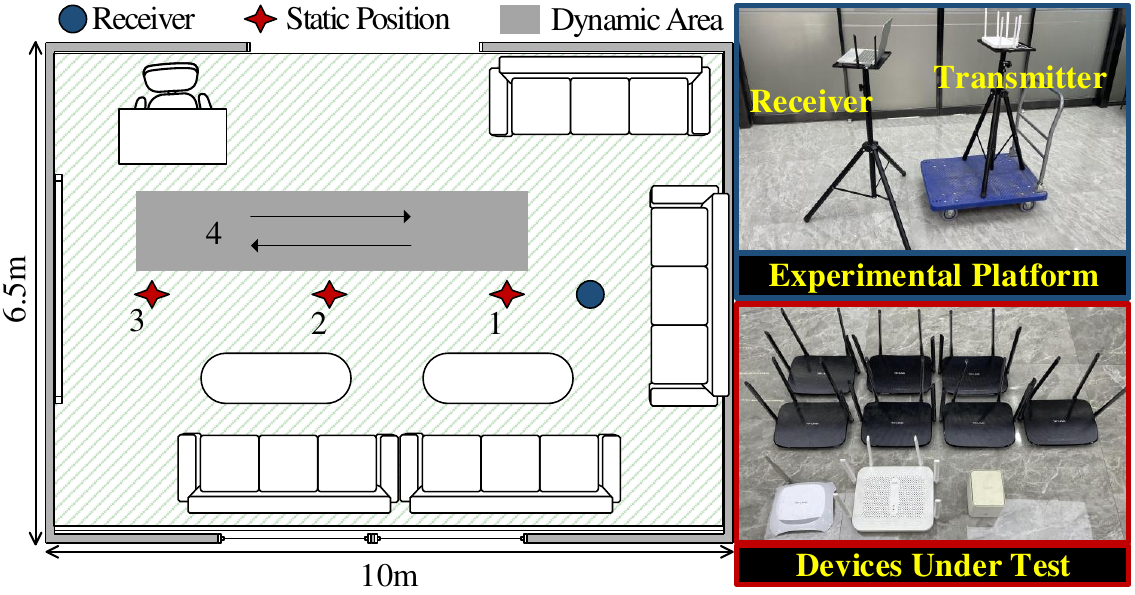}
  \caption{Experimental setting for CSI collection in a controlled scenario.}
  \label{fig:experimental_setting}
\end{figure}

\subsection{Evaluation Methodology}
We build an auxiliary IQ dataset, a synthetic CSI dataset, and two real CSI datasets for model training and testing.

\textbf{Auxiliary IQ Dataset.}
We take raw IQ samples from the open-source WiSig dataset~\cite{11} to obtain our auxiliary dataset.
Specifically, the WiSig dataset consists of transmission signals from 174 commercial WiFi 802.11 devices captured by 41 USRPs.
All devices are set to operate on channel 11 with a center frequency of 2462~MHz and a bandwidth of 20~MHz.
The USRPs capture WiFi signals at a sampling rate of 25~Msps in four different days within a month. 
In this condition, we choose 85 transmitters and 10 receivers and take their IQ samples from the WiSig dataset.
Then, we conduct energy detection, signal segmentation, and resampling to generate 320-point IQ samples.
For each transmitter-receiver pair, we select 30 samples, resulting in a total of 300 samples for each transmitter.
After that, our auxiliary IQ dataset contains more than 25K IQ samples.

\textbf{Synthetic CSI Dataset.}
Based on the auxiliary IQ dataset, we proceed to generate CSI measurements by performing the standard channel estimation on the IQ samples.
To do this, we perform channel estimation using the minimum mean square error method in MATLAB to obtain a dataset that involves about 25K CSI measurements for 85 wireless devices.

\textbf{In-Lab CSI Dataset.}
Besides the synthetic CSI dataset, we collect CSI measurements from ten off-the-shelf WiFi devices in a controlled setting. 
Fig.~\ref{fig:experimental_setting} demonstrates the experimental setting for CSI collection in this scenario.
Specifically, the devices under test include one Honor router, one Redmi router, and eight TP-Link routers.
The receiver is a laptop equipped with an Intel WiFi 6E AX211 network interface card.
The PicoScenes~\cite{31}, a CSI collection tool, is running on the laptop to capture WiFi frames from all routers.
We mount the receiver and transmitters with a height of 1.2~m in a meeting room with a size of 6.5~m $\times$ 10~m.
When collecting CSI measurements, we first place each transmitter 1~m, 3~m, and 5~m away from the receiver, respectively.
Then, we keep the transmitter in a moving state.
In each scenario, we collect 50 CSI measurements from each router, resulting in a total of 200 samples per device.
Finally, we obtain the in-lab CSI dataset with 2K CSI samples.

\begin{figure}[t]
  \centering
  \includegraphics[width=\linewidth]{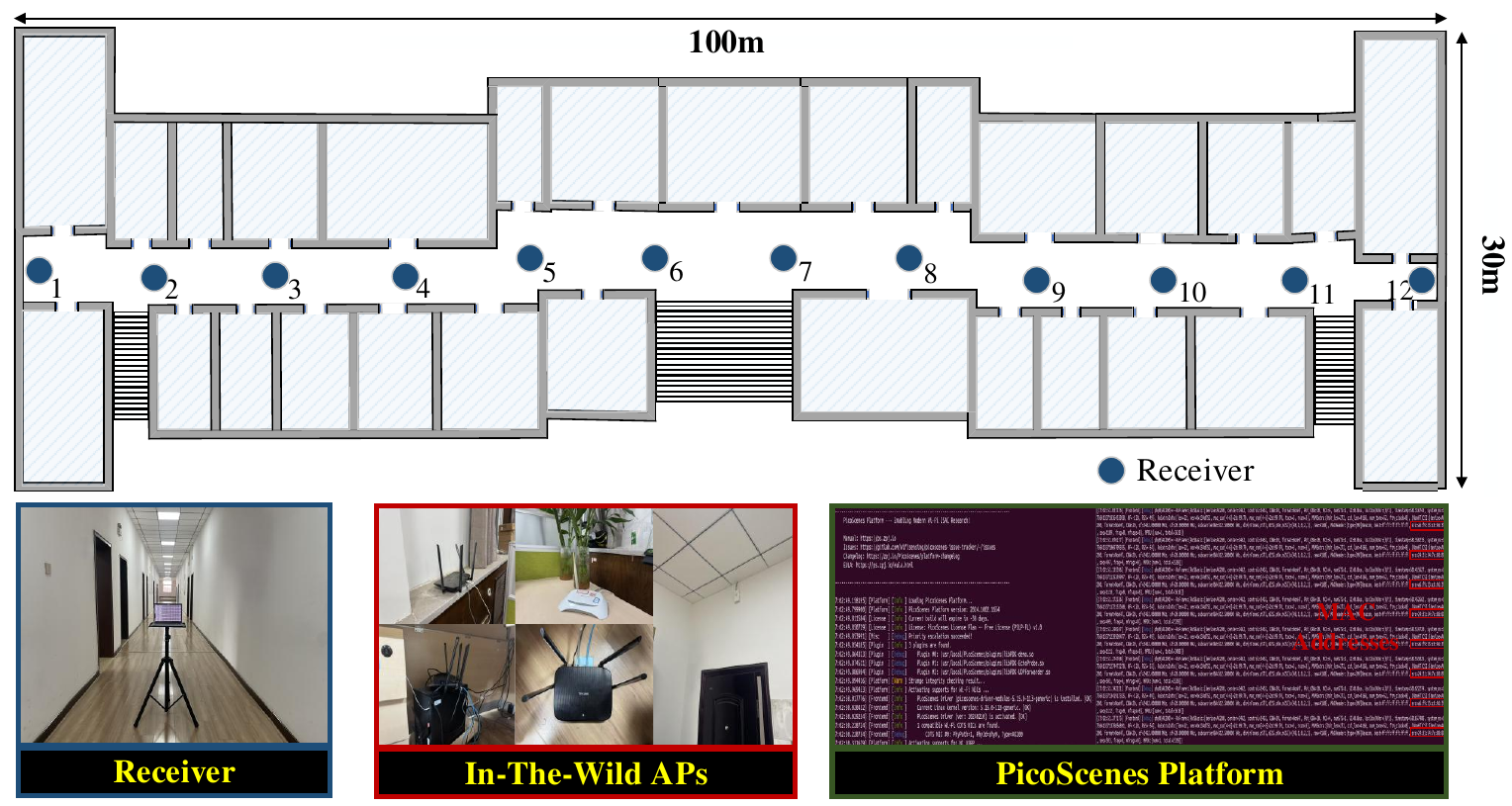}
  \caption{Experimental setting for CSI collection in an uncontrolled scenario.}
  \label{fig:experimental_setting_2}
\end{figure}

\textbf{In-The-Wild CSI Dataset.}
Additionally, we obtain a real CSI dataset in an uncontrolled setting.
To do this, we collect CSI measurements in a building with a size of 100~m $\times$ 30~m on our campus, as illustrated in Fig.~\ref{fig:experimental_setting_2}.
The building layout is complex and uncontrollable human movements exist in this environment.
In this scenario, the transmitters are not deployed by us in advance but are commercial wireless APs used for campus WiFi or private WiFi networks.
The used receiver is consistent with that in collecting the in-lab CSI dataset, and we set the PicoScenes tool to monitor mode to record CSI data from surrounding APs.
In most cases, the transmission between a receiver-transmitter pair is non-line-of-sight.
Since such APs are uncontrollable, we rely on MAC addresses to differentiate them.
Undoubtedly, CSI measurements in this setting are more dynamic and changeable, making it more challenging for CSI fingerprinting in this scenario.
We capture CSI data at 12 receiver locations on the building corridor, and the whole data collection process spans four months.
Finally, we collect 5.18~GB of raw data and select 25 transmitters whose signals appear most frequently.
Table~\ref{tab:Receiver Location–Transmitter Mapping} reports APs present in each receiver location, showing that most APs can be heard in multiple spots. 
We randomly select 1K CSI measurements for each transmitter and obtain the in-the-wild CSI dataset with 25K samples.

\begin{table}[t]
    \centering
    \caption{APs in Each Receiver Location in Uncontrolled Data Collection}
    \label{tab:Receiver Location–Transmitter Mapping}    
    \begin{tabular}{ccc}
      \toprule
       \textbf{Receiver Location} & \textbf{AP ID} \\
      \midrule
       1 & \#7, \#11, \#18 \\
       2 & \#1, \#7, \#11, \#13, \#15, \#18\\
       3 & \#3, \#7, \#8, \#9, \#25\\
       4 & \#8, \#9, \#20, \#25\\
       5 & \#8, \#9, \#20, \#25\\
       6 & \#8, \#9, \#20\\
       7 & \#20, \#24\\
       8 & \#22, \#24\\
       9 & \#4, \#14, \#17, \#24\\
       10 & \#2, \#10, \#14, \#24\\
       11 & \#6, \#10, \#12, \#14, \#16\\
       12 & \#5, \#10, \#16, \#19, \#21, \#23\\
      \bottomrule
    \end{tabular}
\end{table}

\textbf{Implementation.}
After obtaining four datasets, we implement the proposed system in a desktop computer equipped with an Intel Core i5-9400 processor, running the Windows 11 operating system.
We perform the CIM and TDSG processes in MATLAB R2016b, while the neural networks are implemented in PyTorch version 1.13.1, using Python version 3.7. 
We construct different versions of feature extractors using three types of network structures, namely convolutional neural networks (CNNs), recurrent neural networks (RNNs), and temporal convolutional networks (TCNs), each of which has four layers of neurons.
In addition, we build the CSI classifier and the IQ discriminator using two fully connected (FC) layers with a Softmax function to output probability vectors.
The auxiliary network is configured with two FC layers to generate high-dimensional feature vectors.
The Adam optimizer is chosen for all models in the training phase.
When training each model, the initial learning rate is set to 0.001, and the epoch size to 100.
Moreover, the cosine annealing algorithm is utilized to gradually reduce the learning rate.
In addition, we empirically set the hyperparameter $\lambda = 0.30$ and the threshold $\delta = 0.15$ in our evaluation.

\textbf{Evaluation Metrics.} 
We use the following metrics to evaluate the performance of our system. 
\begin{itemize}
    \item \textbf{Accuracy.} It is defined as the ratio of the number of samples that are correctly classified to the total number of samples.
    \item \textbf{F1 Score.} It is the harmonic mean of precision and recall, which comprehensively reflects the model's classification performance.
\end{itemize}

We have made our code and datasets publicly available on GitHub~\cite{dataset}.

\subsection{Evaluation on Synthetic CSI Dataset}

\begin{figure}[t]
  \centering
  \captionsetup[subfigure]{font=scriptsize} %
  \subfloat[]{\includegraphics[width=0.48\linewidth]{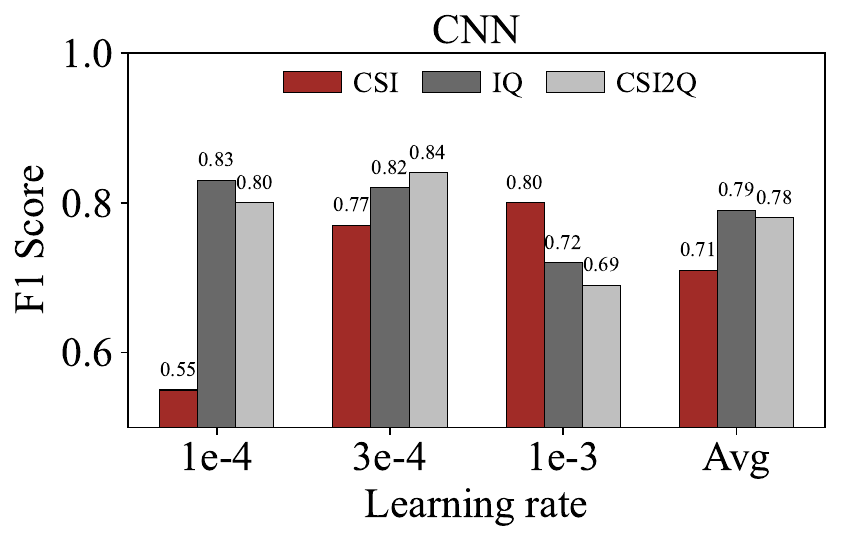}\label{fig:a}}
  \hfill
  \subfloat[]{\includegraphics[width=0.48\linewidth]{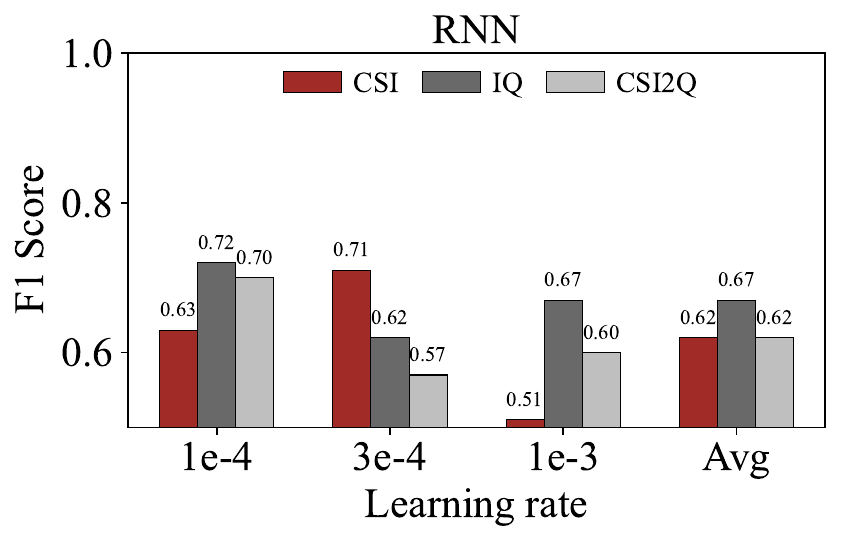}\label{fig:b}}\\[1ex]
  \subfloat[]{\includegraphics[width=0.48\linewidth]{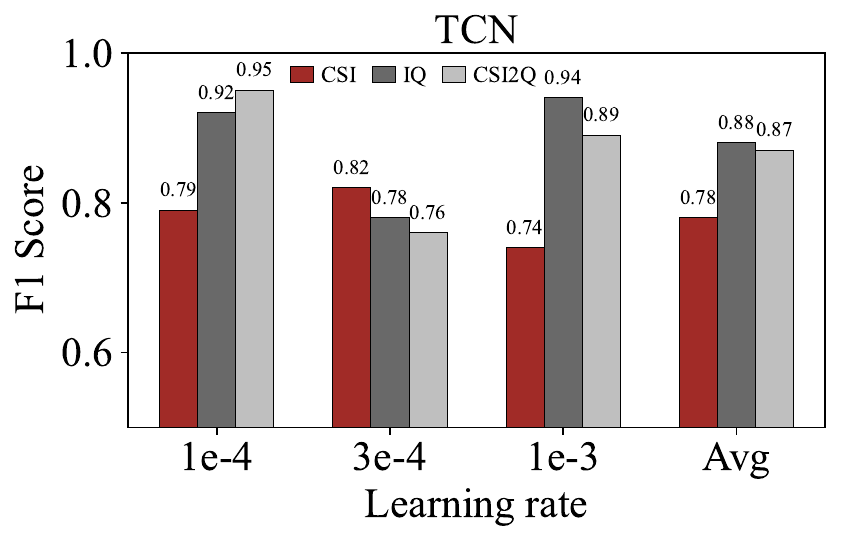}\label{fig:c}}
  \hfill
  \subfloat[]{\includegraphics[width=0.48\linewidth]{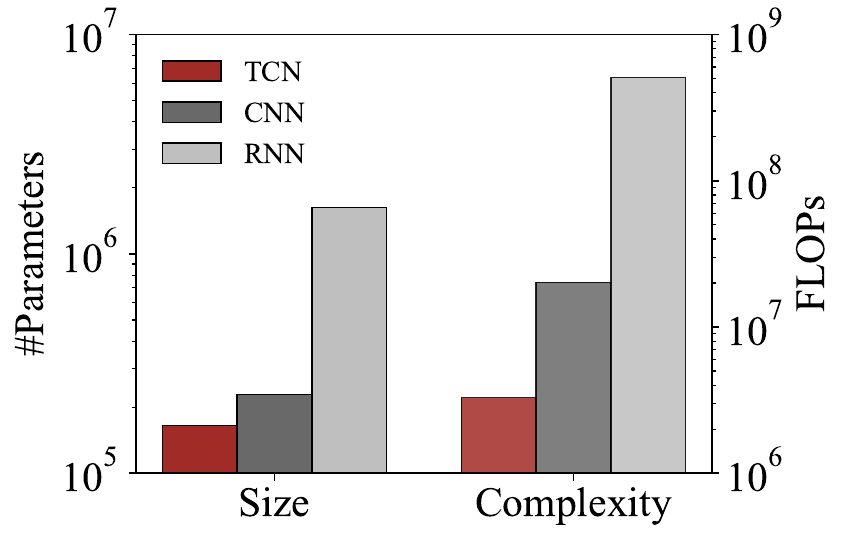}\label{fig:d}}
  \caption{Performance of different models. 
  (a) Classification performance of CNN.
  (b) Classification performance of RNN.
  (c) Classification performance of TCN.
  (d) Network size and computational complexity of three models.}
  \label{fig:model_comparison}
\end{figure}

\textbf{Performance in Closed-World Scenario.}
First, we demonstrate the difference between IQ and CSI data in RF fingerprinting.
For this purpose, we select three candidate models, including a CNN, an RNN, and a TCN, each with two FC layers followed by a Softmax function.
CNN and RNN are two mainstream deep learning networks in current RF fingerprinting solutions~\cite{8,20,33,34,35}.
We directly train three candidate models on the auxiliary IQ dataset and the synthetic CSI dataset, respectively, using three different learning rates, including $1\text{e}^{-4}, 3\text{e}^{-4}, \text{and} \; 1\text{e}^{-3}$.
85 transmitters in the auxiliary and synthetic datasets are used in both the training and testing phases.
For each transmitter, we randomly select 240 samples for training and reserve the remaining 60 samples for testing.
Although raw IQ samples theoretically contain more information than transformed IQ data (CSI\textsuperscript{2}Q) and CSI measurements, not every model achieves the best performance on the IQ data in each setting, as depicted in Fig.~\ref{fig:model_comparison}~(a), (b), and (c).
This is because the performance of deep learning models is highly sensitive to training parameters~\cite{goodfellow2016deep}.
However, on average, each model shows the highest F1 score on the raw IQ samples and has the lowest performance on the CSI data.
Moreover, the transformed IQ data makes the CNN and TCN obtain comparable performance to the IQ samples.

In addition, the TCN achieves the best performance in most cases and shows the highest average F1 scores on different data.
The reason is that the TCN is a deep learning model designed for sequential modeling tasks~\cite{bai2018empirical}.
On the one hand, the dilated convolutions employed by TCNs can model long-range temporal dependencies while mitigating the gradient issues commonly encountered in RNNs. 
On the other hand, compared to CNNs, TCNs provide a larger receptive field with lower complexity.
The above advantages make the TCN a powerful tool in processing sequential data, like IQ samples or CSI measurements.
To verify these, we report the computational overhead of the three models using the number of network parameters and floating point operations (FLOPs) in Fig.~\ref{fig:model_comparison}~(d).
The two metrics of the RNN are at least 10 times higher than those of the TCN and CNN, because the RNN has a recurrent structure that stores and updates hidden states at each time step, leading to a large number of connections and network parameters.
In addition, the TCN presents the fewest network parameters and FLOPs.
This is due to that the TCN uses dilated causal convolutions with a fixed set of convolutional filters and shares parameters across time steps, thus reducing learnable parameters.
Based on the above observations, we adopt the TCN model as our feature extractor with a learning rate of $1\text{e}^{-4}$.
In this setting, CSI\textsuperscript{2}Q improves the F1 score from 0.79 to 0.95.
The result demonstrates the effectiveness of our system in improving CSI fingerprinting.

\begin{table}
    \centering
    \caption{Performance on the Synthetic CSI Dataset in the Open-World Scenario}
    \label{tab:Open-world_recognition_Wisig}    
    \begin{tabular}{ccc}
      \toprule
       Classification Models &Accuracy &F1 Score\\
      \midrule
       TCN & 71.61\%& 0.78\\
       TCN + ALIQ& 78.20\%& 0.82\\
       TCN + ALIQ + OpenMax (CSI\textsuperscript{2}Q) & 94.62\%&0.95\\
      \bottomrule
    \end{tabular}
\end{table}

\textbf{Performance in Open-World Scenario.}
In open-world recognition, we randomly select 45 devices from the synthetic CSI dataset as registered devices and another 5 devices as unknown ones.
For each device, we take 240 CSI samples for training and 60 samples for testing.
During the training phase, we only use the samples from registered devices to train our model.
During the testing phase, the testing samples of both registered and unknown devices are fed into classification models.
In this condition, we build two baseline models.
One is a common TCN, and the other is a TCN with ALIQ but without the OpenMax function.
We compare CSI\textsuperscript{2}Q with the two baselines in the same condition and report the results in Table~\ref{tab:Open-world_recognition_Wisig}.
As the table shows, ALIQ can help the traditional TCN achieve an accuracy improvement from 71.61\% to 78.20\%.
The F1 score also increases from 0.78 to 0.82.
Furthermore, the OpenMax function has led to an accuracy increase of approximately 16.4\% and an F1 score improvement of 0.13.
The above results verify the effectiveness of ALIQ and the OpenMax function feature extraction and recognition in open-world recognition. 

\begin{figure}[!t]
\centering
\captionsetup[subfigure]{font=scriptsize} 
\subfloat[]{\includegraphics[width=0.48\linewidth]{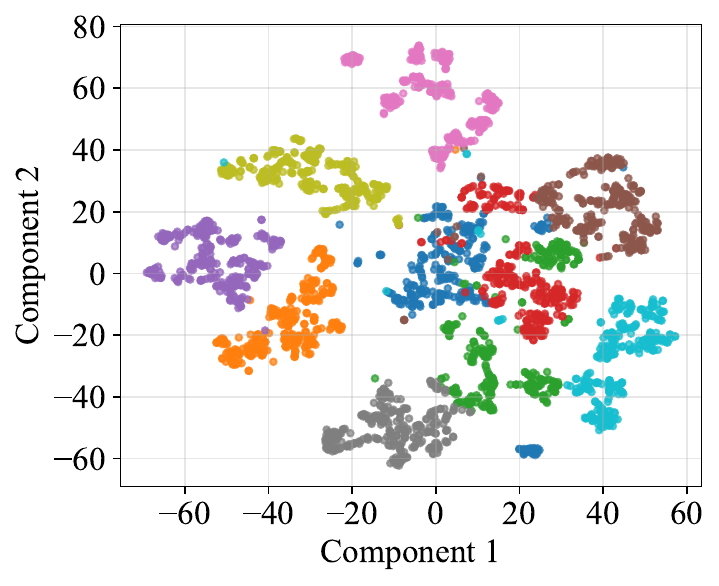}
\label{TDSG_tsne_visualization}}
\hfil
\subfloat[]{\includegraphics[width=0.48\linewidth]{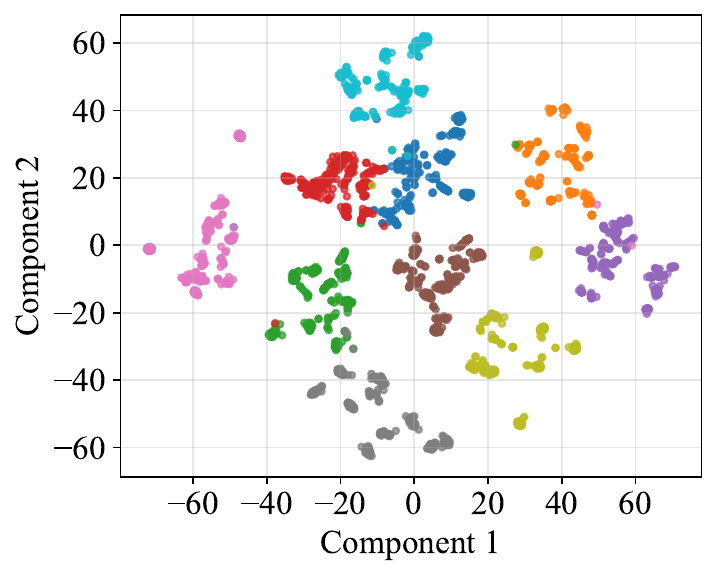}
\label{CIM_TDSG_tsne_visualization}}
\caption{Feature visualization for CSI measurement transformation. (a) TDSG. (b) CIM+TDSG.}
\label{fig:ablation_feature_visualization}
\end{figure}

\textbf{Comparison with Baselines.}
Then, we compare CSI\textsuperscript{2}Q with existing CSI-based device identification approaches. 
Specifically, CSI-RFF~\cite{46} and DeepCRF~\cite{47} are selected as baselines.
We evaluate the three models under both closed-world and open-world scenarios on the synthetic CSI dataset. 
As Table~\ref{tab:Comparison with Baselines} shows, our method consistently achieves the highest accuracy in the two settings.
The reason is that CSI-RFF requires a large number of CSI measurements to generate stable fingerprint features.
When only a limited number of samples are available, the performance of CSI-RFF degrades significantly.
Although DeepCRF also leverages deep neural networks for feature extraction, its generalization capability is limited when a single CSI measurement is used.
By contrast, our approach can extract more robust and distinctive device-specific features from a single CSI measurement via effective transfer learning, demonstrating superior performance in closed-world and open-world scenarios.
\begin{table}
    \centering
    \caption{Comparison with Baselines}
    \label{tab:Comparison with Baselines}    
    \begin{tabular}{ccc}
      \toprule
       Approach & Closed-World Accuracy  &Open-World Accuracy \\
      \midrule
       CSI-RFF~\cite{46}& 74.51\%& 62.80\%\\
       DeepCRF~\cite{47}& 79.49\%& 45.48\%\\
       \textbf{CSI$^{2}$Q (Ours)} & \textbf{95.65\%} & \textbf{94.62\%} \\
      \bottomrule
    \end{tabular}
\end{table}

\textbf{Impact of CSI Measurement Transformation.}
We demonstrate the importance of the CSI measurement transformation component, including the CIM and TDSG methods.
Fig.~\ref{fig:ablation_feature_visualization} illustrates the high-dimensional feature visualization of CSI data after applying our proposed TDSG and CIM+TDSG methods.
Compared to the visualized features in Fig.~\ref{fig:feature visualization}~(b), we can observe that the TDSG scheme reduces the intra-class distances and alleviates the overlap of inter-class feature points as depicted in Fig.~\ref{fig:ablation_feature_visualization}~(a).
Furthermore, when both the CIM and TDSG designs are applied, the features show a great reduction in intra-class distances and a significant increase in inter-class distances as illustrated in Fig.~\ref{fig:ablation_feature_visualization}~(b).
These observations suggest that both the CIM and TDSG procedures of CSI samples are very helpful for CSI fingerprinting. 

\subsection{Evaluation on In-Lab CSI Dataset}

\textbf{Performance in Closed-World Scenario.}
Compared to the synthetic CSI dataset, the in-lab CSI dataset has more complex channel responses.
Specifically, the data in the synthetic CSI dataset only comes from static transmitters, while our CSI data includes measurements from moving transmitters.
Since the devices involved in the in-lab CSI dataset do not have corresponding IQ signals, we still use the auxiliary IQ dataset from the WiSig dataset for auxiliary learning.
In closed-world recognition, the training set comprises 1400 CSI samples from ten devices, while the testing set consists of 600 samples from them.

\begin{figure}[t]
  \centering
  \includegraphics[width=\linewidth]{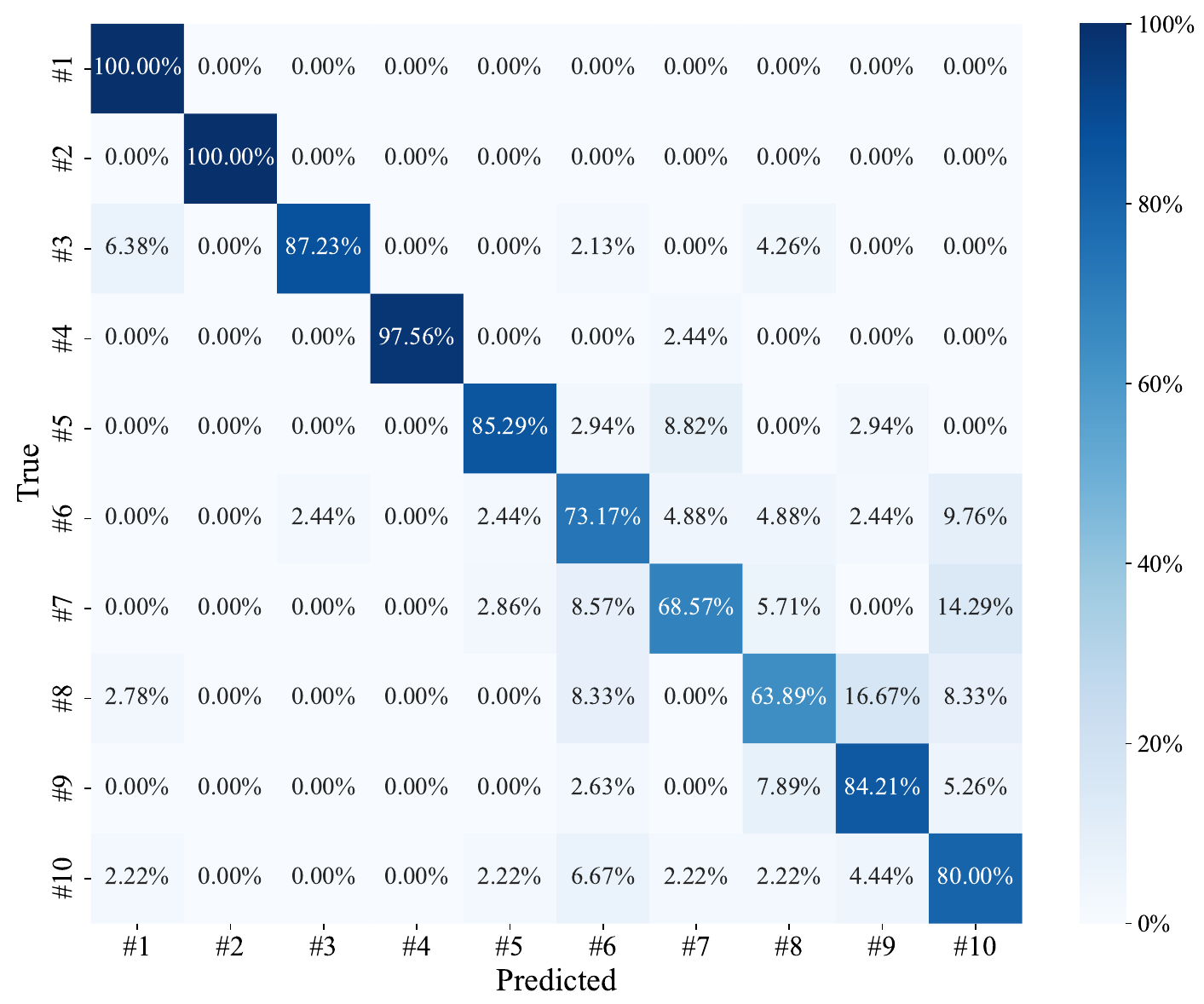}
  \caption{Confusion matrix on the in-lab CSI dataset in the closed-world scenario. }
  \label{fig:confusion_matrix}
\end{figure}

In this setting, we first evaluate our system's performance on each transmitter.
Note that the ten transmitters have different brands.
Therein, the first two represent the Honor and Redmi routers, respectively.
The remaining devices are all TP-Link routers.
The confusion matrix is present in Fig.~\ref{fig:confusion_matrix}.
We can observe that the Honor and Redmi devices exhibit the highest accuracy of 100\% because their brands differ from the others and thus have different RF imperfections, making them easier to discriminate.
In contrast, the seventh and eighth devices have relatively low accuracy of 68\% and 63\%.
The results indicate that the brands of WiFi devices have a significant impact on CSI fingerprinting.

\begin{figure}[t]
  \centering
  \includegraphics[width=1\linewidth]{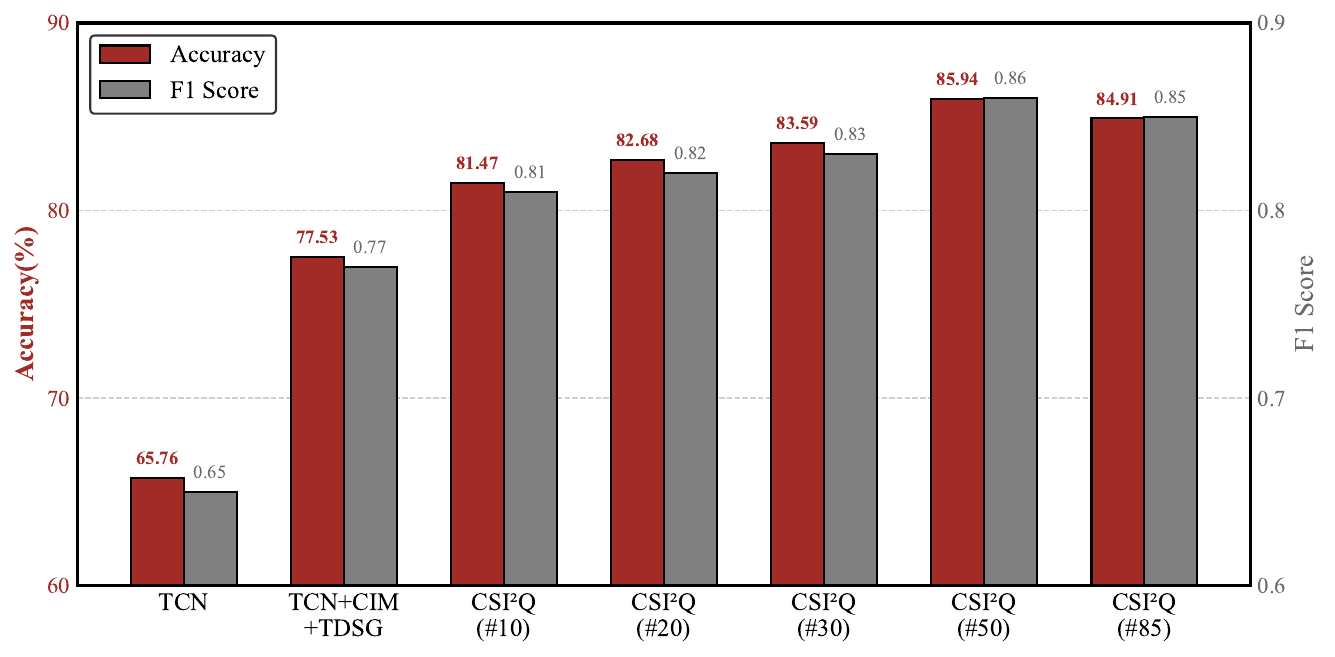}
  \caption{Impact of the device number in the auxiliary dataset.}
  \label{fig:real_csi}
\end{figure}

Then, we investigate the impact of the device number in the auxiliary IQ dataset.
To do this, we vary this number when training the source feature extractor $E_S$ and the IQ discriminator $D$ and record the corresponding results of the CSI classifier.
As illustrated in Fig.~\ref{fig:real_csi}, when only CIM and TDSG are applied, the model accuracy improves from 65.76\% to 77.53\%.
When ALIQ is applied, the classification accuracy can reach 85.94\%.
Thus, even if the IQ samples in the auxiliary dataset do not come from the devices under test, our proposed auxiliary learning approach can still be effective.
Generally, as the auxiliary transmitter number increases, the classification accuracy of the main task also improves.
This is because, when the device number in the source model increases, the model can learn more diverse feature representations.
These learned features can be transferred to the target model by our approach, thereby improving its performance.
However, when the number exceeds 50, the system performance no longer improves.
This may be due to that as the auxiliary device number increases, the common knowledge has been fully learned by the model.
Consequently, the knowledge that can be transferred from the source domain to the target domain is limited.

\begin{table}
    \centering
    \caption{Performance on the In-Lab CSI Dataset in the Open-World Scenario}
    \label{tab:Open-world_recognition_realcsi}    
    \begin{tabular}{ccc}
      \toprule
      Classification Models &Accuracy &F1 Score\\
      \midrule
       TCN & 53.60\%& 0.58\\
       TCN + ALIQ & 65.03\% & 0.65\\
       TCN + ALIQ + OpenMax (CSI\textsuperscript{2}Q) & 81.53\%&0.76\\
      \bottomrule
    \end{tabular}
\end{table}

\textbf{Performance in Open-World Scenario.}
We also conduct an open-world recognition experiment on the in-lab CSI dataset.
We select 8 devices as registered devices and 2 devices as unknown ones.
For each device, we randomly select 160 samples for training and 40 samples for testing.
The training and testing settings are the same as those used in the open-world recognition on the synthetic CSI dataset.
Table~\ref{tab:Open-world_recognition_realcsi} shows the results of our open-world recognition.
CSI\textsuperscript{2}Q achieves an accuracy improvement from 53.60\% to 81.53\% on the in-lab CSI dataset.
The F1 score increases from 0.58 to 0.76.
Additionally, the OpenMax function has led to an accuracy increase of approximately 16.5\% and an F1 score improvement of 0.11.
The above results tell that CSI\textsuperscript{2}Q is effective in handling real CSI measurements in open-world RF fingerprinting.

\subsection{Evaluation on In-The-Wild CSI Dataset}

\begin{figure}[t]
  \centering
  \includegraphics[width=0.92\linewidth]{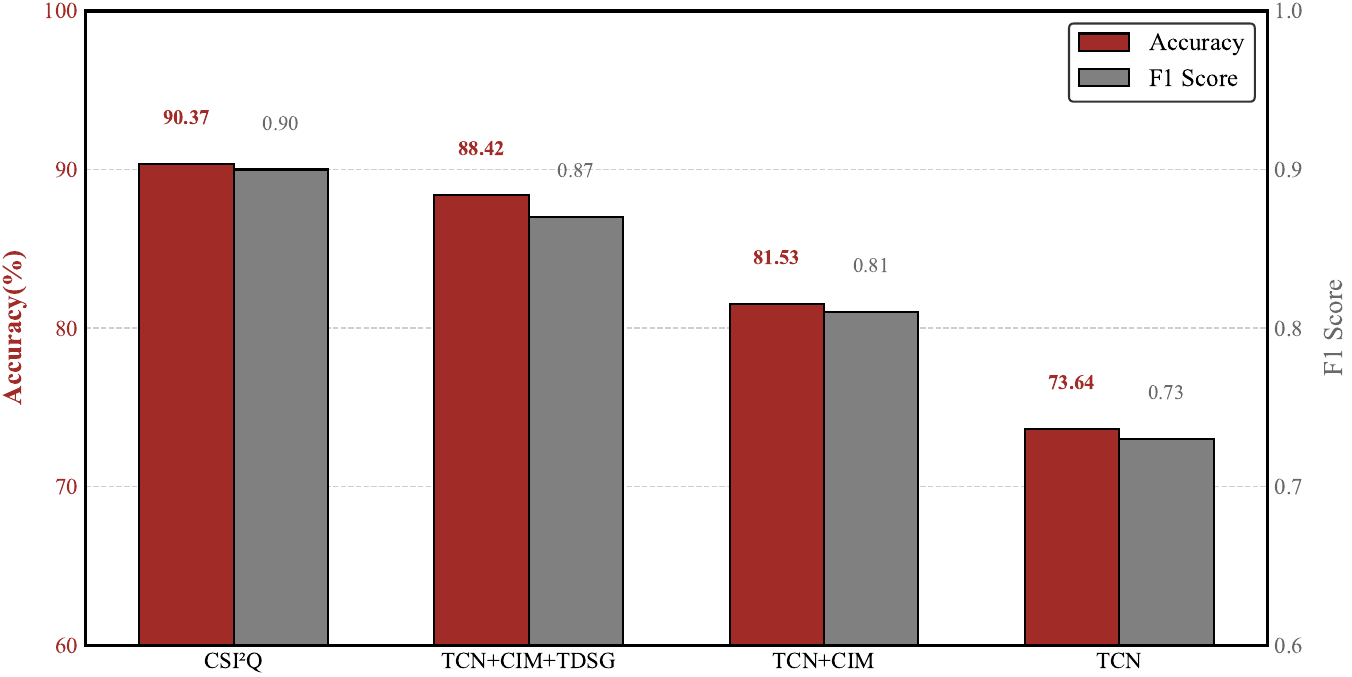}
  \caption{Ablation study of CSI\textsuperscript{2}Q.}
  \label{fig:tcn_ablation}
\end{figure}

\textbf{Ablation Study in Closed-World Scenario.}
Since the in-the-wild CSI dataset is collected in an uncontrollable setting, the CSI measurements have more complex multi-path effects compared to those of the in-lab dataset.
Similarly, the involved devices do not have paired IQ signals, so we still use the auxiliary IQ dataset from the WiSig dataset for knowledge transfer.
In closed-world recognition, the training set comprises 20K CSI samples from 25 devices, while the testing set consists of 5K samples from them.
We perform an ablation study to measure the significance of each component in our system. 
For this purpose, we divide CSI\textsuperscript{2}Q into three parts, i.e., CIM, TDSG, and ALIQ.
We remove the three components one by one, train the remaining parts, and compare their impacts on the system performance. 
As shown in Fig.~\ref{fig:tcn_ablation}, each component of our design has a positive impact on the overall performance.
The combination of CIM, TDSG, and ALIQ achieves a classification accuracy of 90.37\%.
When removing ALIQ, the accuracy decreases to 88.42\%.
When the system consists only of CIM, the accuracy further decreases to 81.53\%.
When the CIM component is also removed, the accuracy drops to only 73.64\%.
A similar trend can also be observed in the F1 score.
The results of our ablation experiments demonstrate that all system designs have a significant impact on the overall performance.
Moreover, these experimental findings further affirm the effectiveness of CSI\textsuperscript{2}Q in improving CSI fingerprinting.

\begin{table}
    \centering
    \caption{Performance on the In-the-Wild CSI Dataset in the Open-World Scenario}
    \label{tab:Open-world_recognition_realcsi_2}    
    \begin{tabular}{ccc}
      \toprule
       Classification Models &Accuracy &F1 Score\\
      \midrule
       TCN & 56.89\%& 0.63\\
       TCN + ALIQ & 72.31\% & 0.74\\
       TCN + ALIQ + OpenMax (CSI\textsuperscript{2}Q) & 82.29\%&0.80\\
      \bottomrule
    \end{tabular}
\end{table}

\begin{figure}[t]
  \centering
  \includegraphics[width=\linewidth]{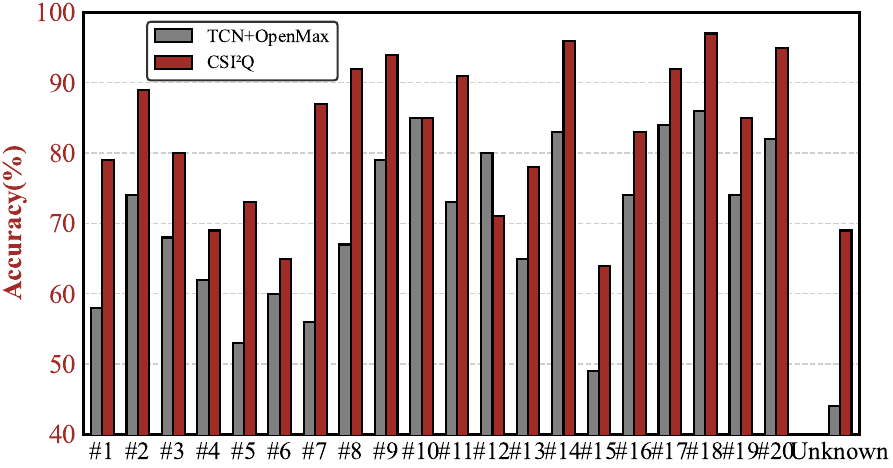}
  \caption{Performance of each device on in-the-wild CSI dataset.}
  \label{fig:Open-world_recognition_realcsi_2}
\end{figure}

\textbf{Performance in Open-World Scenario.}
In open-world recognition, we select 20 devices from the in-the-wild CSI dataset as registered devices and another 5 devices as unknown ones.
For each device, we take 800 CSI samples for training and 200 samples for testing.
During the training phase, we only use the samples from registered devices to train our model.
During the testing phase, the testing samples of both registered and unknown devices are fed into the model.
Table~\ref{tab:Open-world_recognition_realcsi_2} reports the system performance in this setting.
With the help of our system, the TCN achieves an accuracy improvement in open-world recognition from 56.89\% to 82.29\%.
The F1 score also increases from 0.63 to 0.80.
Additionally, the usage of the OpenMax function has led to an accuracy increase of approximately 10\% and an improvement of 0.06 in the F1 score.
Fig.~\ref{fig:Open-world_recognition_realcsi_2} compares the recognition accuracy before and after implementing our proposed system.
We can observe that the recognition accuracy has improved for all devices with the help of CSI\textsuperscript{2}Q.
The recognition accuracy of all devices remains above 64\%, with the best-performing device achieving an accuracy as high as 98\%.
The recognition accuracy for unknown devices ranges from 44.73\% to 69.37\%.
The above results show the effectiveness of our system in open-world recognition. 

\begin{figure}[t]
  \centering
  \captionsetup[subfigure]{font=scriptsize} %
  \subfloat[]{\includegraphics[width=0.48\linewidth]{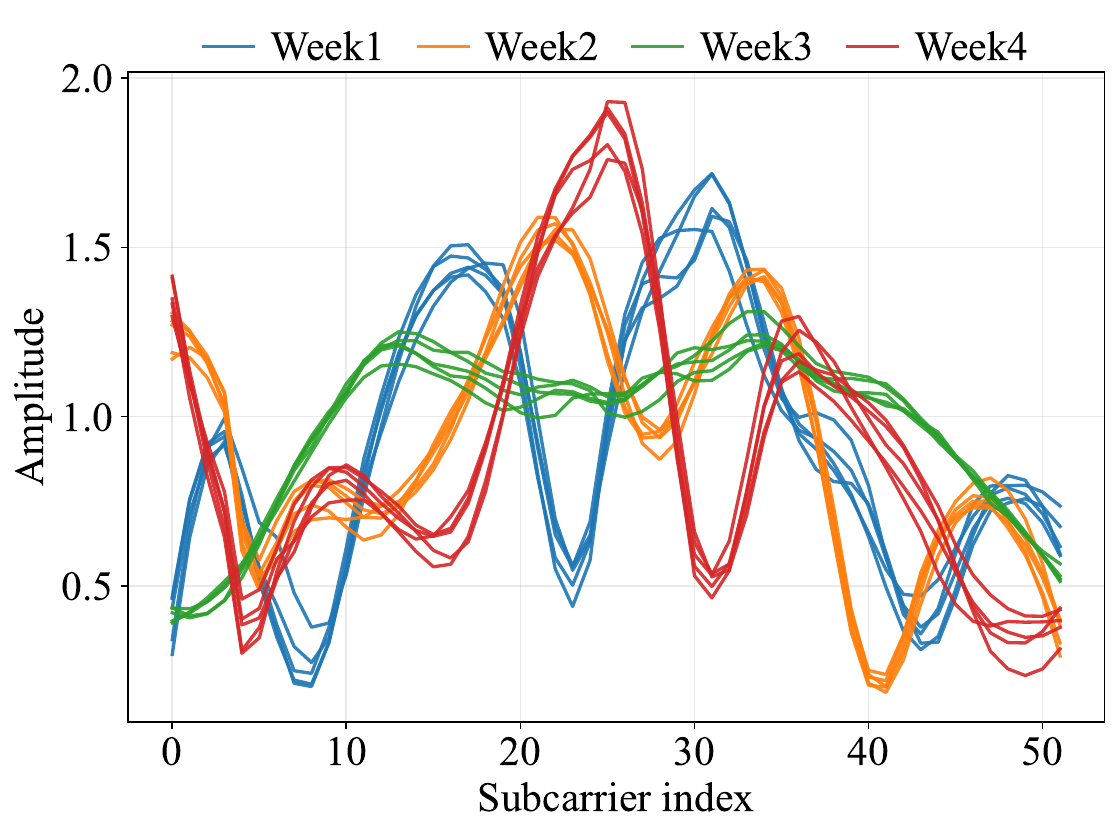}}\hfill
  \subfloat[]{\includegraphics[width=0.47\linewidth]{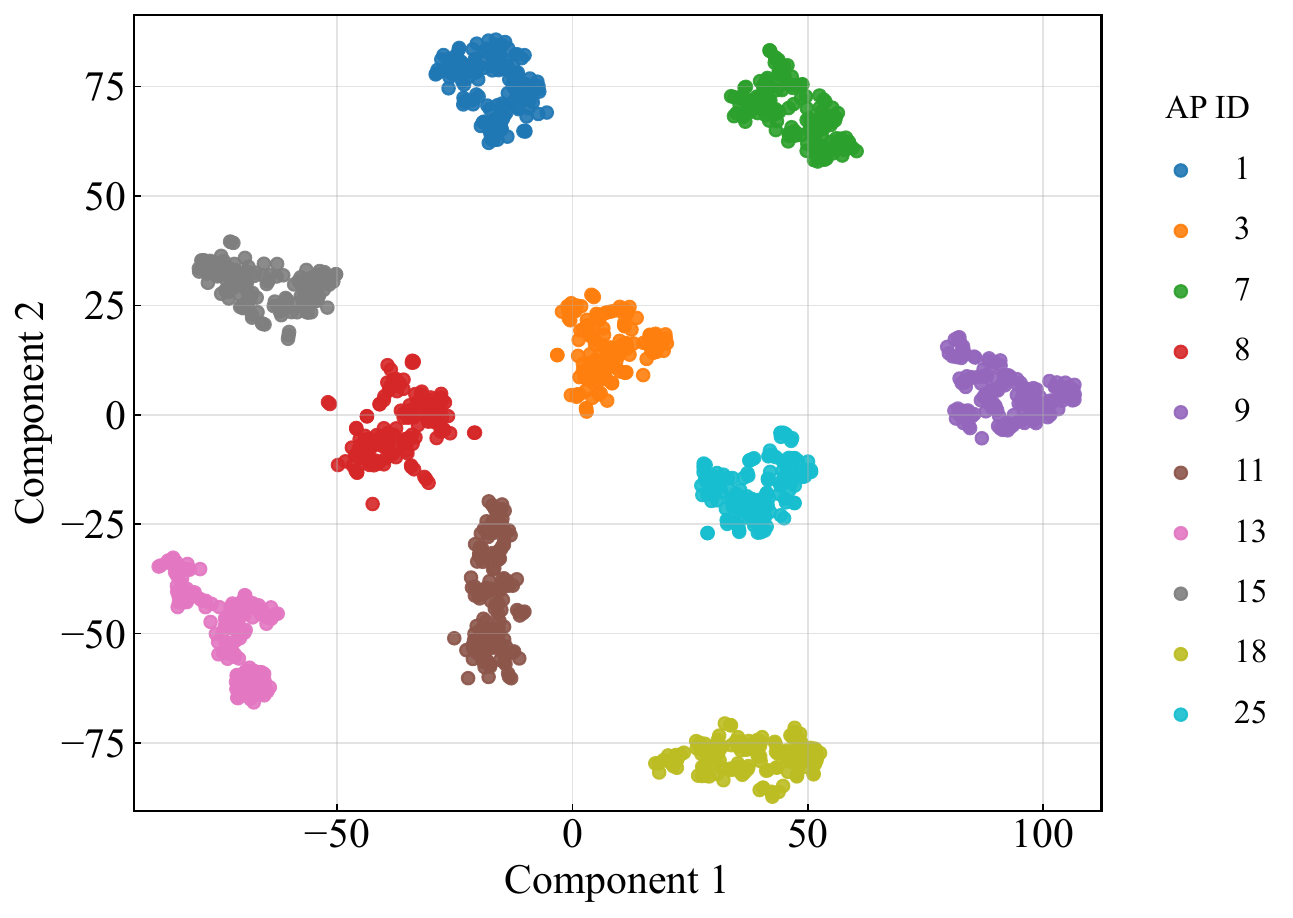}}
  \caption{Impact of receiver locations. 
  (a) CSI measurements of No.1 AP in four consecutive weeks. 
  (b) t-SNE visualization on CSI measurements from the first three locations.} 
  \label{fig:tsne_location}
\end{figure}

\textbf{Impact of Receiver Locations.}
Although some APs appear only once at specific receiver locations as reported in Table~\ref{tab:Receiver Location–Transmitter Mapping}, it is unlikely for our system to associate location-specific channel effects with device identities.
Because the wireless channel changes dramatically in reality, it is improbable that the high consistency of the wireless channel exists in the in-the-wild CSI dataset.
Specifically, we collect CSI measurements once a week for a total of four months.
Although the locations of the 12 receiver points are roughly marked, the deployment position of the receiver and the orientation of its antenna will not be the same in each collection.
Moreover, all APs are uncontrollable and their locations and the surrounding reflective items could alter over four months, making their channel conditions different in multiple collections.
To illustrate this, we present the CSI measurements of the No.1 AP, which only appears at the second receiver location, for four consecutive weeks in Fig.~\ref{fig:tsne_location}~(a).
The CSI measurements are significantly different across four collections, indicating that the signal propagation condition of the No.1 AP is constantly changing.
Furthermore, we perform the t-SNE visualization on the CSI measurements from the first three locations.
The measurements come from ten different APs.
As depicted in Fig.~\ref{fig:tsne_location}~(b), each AP has a clear feature cluster in the latent feature space.
Moreover, all clusters from the three locations are distributed uniformly and in a mixed way in the feature space.
The above observations suggest that our system is unlikely to associate location-specific features with device identities.

\section{Related Work}

\textbf{IQ-Based Approaches.}
For IQ-based RF fingerprinting, Candore et al.~\cite{13} extract CFOs, carrier phase offsets, I/Q phase offsets, and error vector magnitude to measure the differences in devices such as the transmitter antenna and oscillator.
Knox et al.~\cite{14} propose using a noise filter to select the collected signals and then extracting the phase information of the demodulated signals as RF fingerprints, demonstrating their classification performance at different distances.
The work~\cite{15} mathematically models features such as carrier frequency offsets, phase noise, and IQ imbalance of the transmitter, estimating these features using gradient descent algorithms with high accuracy.
Wong et al.~\cite{16} achieve transmitter classification for QAM and PSK modulation schemes by inputting raw IQ samples into a designed CNN IQ imbalance estimator.
Sankhe et al.~\cite{17} propose a new system called ORACLE based on convolutional neural networks to identify wireless devices by learning the fine-grained distortions imposed by the physical layer on IQ samples.
Al-Shawabka et al.~\cite{18} evaluate the impact of wireless channels on classification accuracy using a large-scale dataset and discover that equalizing the IQ data can improve classification accuracy.
Elmaghbub et al.~\cite{19} study the impact of dynamic channels on LoRa RF fingerprint recognition and reveal the sensitivity of deep learning-based fingerprint recognition schemes to various channel variations.
Soltani et al.~\cite{20} explore methods to improve RF fingerprint classification accuracy using data augmentation, introducing interference from channel and noise into the IQ data training set.
Most IQ-based RF fingerprinting approaches achieve high recognition performance.
However, costly and dedicated devices like USRPs are required to capture raw IQ signals, which hinders their practical application in the real world.

\textbf{CSI-Based Approaches.}
In recent years, some researchers have explored the utilization of CSI measurements for RF fingerprinting.
Hua et al.~\cite{21} are the first researchers who extract carrier frequency offsets from CSI measurements for detecting illegal devices.
The work~\cite{22} extracts the phase errors that are robust to time variations and position changes for user authentication.
DeepCSI~\cite{23} is proposed to authenticate WiFi devices based on CSI feedback.
R-RRF~\cite{24} also uses CSI feedback to eliminate the influence of multipath propagation by establishing the RF impairment relationship between antenna chains of multi-antenna devices.
The literature~\cite{25} proposes micro-CSI that extracts the channel-independent features under the line-of-sight (LoS) scenarios to identify devices.
More recently, CSI-RFF~\cite{46} employs a signal space-based method to extract fine-grained fingerprints caused by RF hardware imperfections under LoS conditions. 
To improve robustness under non-LoS conditions, DeepCRF~\cite{47} extends this approach with a deep learning method, using contrastive learning and data augmentation.
However, existing CSI fingerprinting systems only focus on how to eliminate the influence of wireless channels in CSI measurements, without considering the issues of coarse granularity in open-world scenarios.

\section{Conclusion}
This paper presents CSI\textsuperscript{2}Q to achieve CSI fingerprinting performance comparable to IQ-based approaches.
We observe that CSI and IQ data are inherently correlated, and CSI fingerprinting can benefit from IQ samples.
Instead of extracting RF fingerprints directly from raw CSI measurements, CSI\textsuperscript{2}Q transforms them into time-domain signals that share the same feature space with IQ samples. 
Then, the strong feature extraction ability of an IQ fingerprinting model is transferred to its CSI counterpart via an auxiliary training strategy.
Finally, the trained CSI fingerprinting model is combined with an OpenMax function to calibrate the uncertainty of registered devices and thus estimate the likelihood of unknown ones.
We evaluate CSI\textsuperscript{2}Q on both synthetic and real CSI datasets in closed- and open-world settings.
On the synthetic dataset, our system achieves an accuracy improvement of about 16\% in the closed-world scenario and about 23\% in the open-world case.
On the in-lab dataset, it obtains an accuracy increase of about 20\% in the closed-world scenario and about 28\% in the open-world setting.
On the in-the-wild dataset, our system achieves a 17\% increase in accuracy in the closed-world scenario and a 25\% increase in accuracy in the open-world scenario.

\bibliographystyle{IEEEtran}
\bibliography{references}
\vfill

\end{document}